\documentclass{aa}  
\usepackage{natbib}
\bibpunct{(}{)}{;}{a}{}{,}

\usepackage{graphicx}
\usepackage{txfonts}
\usepackage{hyperref}

\usepackage{caption}
\usepackage{subcaption}
\usepackage{lscape}
\usepackage{gensymb}
\usepackage{float}

\usepackage{xcolor}

\newcommand*{\teff}{$T_\mathrm{eff}$}

\newcommand*{\kms}{km~s$^{-1}$}
\newcommand*{\masyr}{mas~yr$^{-1}$}

\begin{document}

   \title{Painting a Family Portrait of the Yellow Super- and Hypergiants in the Milky Way}

   \subtitle{I. Constraining the Distances and Luminosities}

   \author{A.~Kasikov \inst{\ref{to},\ref{eso}} \and A.~Mehner \inst{\ref{eso}} \and I.~Kolka \inst{\ref{to}} \and A.~Aret \inst{\ref{to}}}

   \institute{Tartu Observatory, University of Tartu, Observatooriumi 1, Tõravere, 61602, Estonia\label{to}\\
    \email{anni.kasikov@ut.ee}
    \and
    European Southern Observatory, Alonso de C\'ordova 3107, Vitacura, Casilla 19001, Santiago de Chile, Chile\label{eso}
    }

   \date{Received ; accepted }

  \abstract
   {Distances to evolved massive stars in the Milky Way are not well constrained by \textit{Gaia} parallaxes due to their brightness and variability. This makes it difficult to determine their fundamental stellar parameters, such as radius or luminosity, and infer their evolutionary states.}
   {We aim to improve the distance estimates of Yellow Hypergiants (YHGs) and Yellow Supergiants (YSGs) by identifying possible cluster and association memberships. Using these distances, we derived updated luminosities and revised their positions in the Hertzsprung-Russell diagram. }
   {We compiled from the literature a sample of 35 luminous yellow massive stars (YHGs and the most luminous YSGs). We used \textit{Gaia} DR3 astrometry to identify possible membership in clusters and OB associations. We derived distances by combining the parallaxes of nearby co-moving stars. We independently validated these distances by comparing the stellar radial velocities to the Galactic \ion{H}{I} kinematic map. We combined angular diameters and effective temperature values from the literature with the new distances to estimate luminosities. }
   {We improved the distance estimates for 28 of the 35 stars through association with co-moving stellar groups. For an additional six stars, we provided distance estimates based on the \ion{H}{I} kinematic map. For one star, the distance remains unclear. Most YSGs are members of young stellar populations, while the environments of the YHGs are more diverse, and for some of them their origin populations remain unclear. We derived updated luminosities for a subset of 20 stars. Most YHGs have luminosities above $\log L/L_\odot=5.4$, while YSGs occupy a wider range of luminosities and the most luminous YSGs have luminosities similar to YHGs. }
   {}


   \maketitle
%
%-------------------------------------------------------------------

\section{Introduction}

Stellar population studies of young star clusters have revealed numerous massive stars in transitional evolutionary phases \citep[recently, e.g.][]{marco_ngc_2025,maiz_apellaniz_barba_2025}. Knowledge of the surrounding stellar population and environment provides context for understanding the evolution of massive stars. Yellow Supergiants (YSGs) and the rarer, more elusive Yellow Hypergiants (YHGs) are unstable and short-lived phases on the cool side of the upper Hertzsprung-Russell (HR) diagram. YHGs are generally considered to represent a more extreme or advanced stage of evolution than typical YSGs and have been proposed as post-Red Supergiants (post-RSGs) \citep{de_jager_yellow_1998}. 

The distinction between YHGs and YSGs is based on their spectroscopic characteristics: YHGs have extended atmospheres, one or more broad H$\alpha$ emission components, and significantly wider absorption lines compared to YSGs \citep{de_jager_yellow_1998}. Beyond using spectroscopic features to differentiate YHGs from YSGs, studies have proposed luminosity thresholds \citep[$\log L/L_\odot>5.2$;][]{de_jager_obstacle_1997}, distinct pulsational behaviour and presence of circumstellar dust \citep{humphreys_yellow_2023}, and a high \element[ ][12]{CO}/\element[ ][13]{CO} isotopic ratio \citep{oksala_probing_2013}. 
Homogeneously determined observational parameters of YHGs and YSGs provide a basis for comparison with stellar evolutionary models, improving our overall understanding of evolutionary pathways of luminous yellow massive stars and their role as supernova progenitors \citep[e.g.][]{aldering_sn_1994,crockett_type_2008,maund_yellow_2011,tartaglia_progenitor_2017,kilpatrick_progenitor_2017,niu_discovery_2024,reguitti_sn_2025}.

Extensive work has been done to characterise the luminous cool Supergiant populations of the Magellanic Clouds and other nearby galaxies \citep[e.g.][]{martin_census_2023,dorn-wallenstein_physical_2023,maravelias_machine-learning_2025}. 
However, in the Milky Way, studies have been hampered by uncertain distances and high extinction in the Galactic plane.  
In the era of \textit{Gaia} and due to the efforts by \citet{bailer-jones_estimating_2021}, the distance estimates for many stars have improved significantly. For objects that lack reliable \textit{Gaia} parallaxes, other indirect methods can be used to estimate distances. We combined and compared the results of two complementary methods. The first method is based on analysis of nearby stars: since many YHGs and YSGs are known or suspected members of stellar clusters or OB associations, confirming these memberships and combining the parallaxes of nearby stars \citep[e.g.][]{campillay_spectroscopic_2019} can help refine their distance estimates. The second method compares the stellar radial velocities with the Galactic \ion{H}{I} kinematics. Agreement between these independent methods increases confidence in the derived distances.  

In Sect.~\ref{sect:sample}, we give an overview of the selected targets. In Sect.~\ref{sect:distance}, we revisit the cluster and OB association memberships of YHGs and YSGs in the Milky Way and determine the distances to the stellar groups. In Sect.~\ref{sect:hi_dist}, we derive independent distance estimates using the \ion{H}{I} kinematic map of the Milky Way. In Sect.~\ref{sect:rad_lum}, we combine our distance measurements with effective temperatures and angular radii from the literature to place the Milky Way YHGs and YSGs on the HR diagram. Finally, in Sect.~\ref{sect:discussion}, we discuss the spatial distribution and binary properties of our sample stars.

\section{Sample selection}\label{sect:sample}

We queried Simbad for YSGs and F- and G- spectral class stars with luminosity class Ia or Iab with infrared excess ($V-K > 3$~mag). From the resulting list, we excluded objects that are identified in the literature as luminous post-asymptotic giant branch (post-AGB) stars. We also excluded stars with reliable \textit{Gaia} DR3 parallaxes that place them closer than expected for being luminous YSGs. We added additional luminous YSGs from the literature \citep{de_jager_yellow_1998,mantegazza_luminosities_1992,kovtyukh_accurate_2012} and included stars that have been proposed as YHG candidates: \object{IRAS 18357-0604} \citep{clark_iras_2014}, and \object{HD 144812} \citep{kourniotis_hd_2025}. The resulting 25 stars form our sample of YSGs. 

We also included the well-known YHGs in the Milky Way: \object{V509 Cas}, \object{$\rho$ Cas}, \object{IRC +10420}, \object{HR 5171}, \object{6 Cas}, \object{HD 96918}, and the Yellow-Red Hypergiant \object{RW Cep} \citep{de_jager_yellow_1998}; \object{IRAS 17163-3907} \citep{lagadec_double_2011};  \object{HD 179821} \citep{hawkins_discovery_1995}; and \object{[FMR2006] 15} \citep{figer_discovery_2006}. When mentioning YHGs throughout this paper, we refer to the ten stars in this list. 

We are not aiming for a complete census, but rather for a representative sample of the most luminous YSGs and YHGs. The final sample of 35 objects is presented in Table~\ref{tab:distances}.  

\section{Distances based on stellar group identification}\label{sect:distance}

\textit{Gaia} Data Release 3 (DR3) \citep{gaia_collaboration_gaia_2016,gaia_collaboration_gaia_2023} has provided proper motion and parallax measurements for $\sim$1.3 billion sources, enabling improved determinations of distances and cluster memberships. However, \textit{Gaia} parallaxes are subject to systematics and biases. For YHGs and YSGs, the parallax values can have large uncertainties (>20\%) and, in some cases, they are even negative. A major source of error is their brightness; approximately half of our sample is brighter than $G = 6$~mag. Another source of astrometric error for cool massive stars is caused by photocentric variability due to surface convection \citep{chiavassa_radiative_2011,pasquato_limits_2011,el-badry_how_2025}. Additionally, some stars have high values of renormalised unit weight error (\textsc{ruwe}), indicating poor astrometric fits, which may result from an unresolved binary companion \citep[e.g.][]{castro-ginard_gaia_2024} or circumstellar structure \citep{fitton_disk_2022}.

\cite{bailer-jones_estimating_2021} provided probabilistic distance estimates that take into account the Galactic structure and mitigate the limitations of the simple 1/parallax distance estimate. Although generally more reliable, these distance estimates still require careful interpretation. In several cases, the \citet{bailer-jones_estimating_2021} distances disagree with the values commonly adopted in the literature (see Table~\ref{tab:distances}). For stars with \textsc{ruwe}$<1.4$ and parallax uncertainties better than 20\%, the \citet{bailer-jones_estimating_2021} distances are likely to be reliable.

To improve the distances for YHGs and YSGs with poor \textit{Gaia} parallaxes, we explored their kinematics based on proper motions in the context of their surrounding stellar environments -- nearby star clusters, OB associations, and regions hosting Young Stellar Objects (YSOs). We cross-matched several catalogues of open clusters and OB associations with \textit{Gaia} data. Membership in such co-moving groups implies common distance. 

\subsection{Method}

To determine possible open cluster memberships, we queried the catalogue of \citet{hunt_improving_2024}. We counted a YHG/YSG as a member of a cluster if its projected on-sky position is within the cluster boundaries and its proper motion is within 3$\sigma$ of the cluster's mean proper motion. When reliable parallaxes are available, we also used them to confirm membership. 
We checked whether the stars are members of any known OB associations listed by \citet{melnik_kinematics_2017}. We compared the proper motions of the YHGs and YSGs with members of the OB associations in the catalogue by \cite{chemel_search_2022} to infer a possible affiliation. OB associations extend farther than clusters in the area projected on sky and are less tightly connected in proper motion space. 
We counted a YHG/YSG as affiliated with an OB association when at least 2 association members are located within a 1~deg radius on the sky and within a 0.5~mas~yr$^{-1}$ radius in proper motion space. In most cases, there are significantly more stars with similar proper motion in the same sky region. 
We list possible cluster and OB association memberships in the "Identified cl/assoc" column of Table~\ref{tab:distances}. 
Detailed notes for all individual stars are provided in App.~\ref{sect:distance_comments_stars}. 

To estimate the group-based distance for each target, we identified stars with high-quality astrometric data belonging to the same population. Proper motions and positions are not enough to differentiate between stellar populations, another criterion is needed \citep[e.g. colour index was used by][]{negueruela_westerlund_2022}. We used the effective temperature $T_\mathrm{eff}$ from \textit{Gaia} GSP-Phot as an additional criterion. Since we searched for co-located and co-moving populations associated with young clusters or OB associations, we limited our search to B-type and early A-type stars with effective temperatures $8\,700\mathrm{K}<T_\mathrm{eff}<18\,000\mathrm{K}$. This approach assumes that YSGs/YHGs are still physically associated with their birth environments. If the star has migrated or has been ejected from its birth environment, we might not be able to identify a co-moving group, and this method becomes unreliable. We followed the methodology given by \citet{campillay_spectroscopic_2019} and \citet{maiz_apellaniz_lucky_2021,maiz_apellaniz_validation_2021}. 

We identified hot stars belonging to the neighbourhood of each YHG and YSG, combined their parallaxes, and estimated the distance to the stellar group following these steps:
\begin{enumerate}
    \item Select hot stars with high-quality astrometry within 10{\arcmin} of the target, applying the following selection criteria: good astrometry (\textsc{ruwe}$<1.4$); bright sources ($G<18$~mag); reliable parallaxes (parallax/error > 5); 5-parameter astrometric solution (\textsc{astrometric\_params\_solved}=31). 
    \item Correct the \textit{Gaia} proper motions for known biases affecting bright targets, following the methodology of \cite{cantat-gaudin_characterizing_2021}.
    \item Apply a proper motion cut to separate stars with similar kinematics.
    \item Correct the parallaxes for \textit{Gaia} zero point offsets \citep{lindegren_gaia_2021}\footnote{\raggedright Using the Python code provided at \url{https://gitlab.com/icc-ub/public/gaiadr3_zeropoint}}.
    \item Calculate the group parallax, combining the individual parallaxes and errors following the recipe given in \citet{campillay_spectroscopic_2019}, considering the external parallax uncertainty \citep{fabricius_gaia_2021} and the angular covariance term. The systematic parallax uncertainty gives a minimum distance uncertainty of $\approx d$\% for a star at d~kpc \citep{maiz_apellaniz_validation_2021}.
    \item Convert the group parallax into a geometric distance, using the generalized gamma distribution (GGD) prior of \citet{bailer-jones_estimating_2021}\footnote{\raggedright Using the interactive Jupyter Notebook based tool \url{https://github.com/ElisaHaas25/Interactive-Distance-Estimation}}
\end{enumerate}

The range of the proper motion cut is flexible and depends on the stellar environment of each YSG in our sample. By exploring the scatter of proper motions of stars recognised as members of a single cluster in the catalogue of \citet{hunt_improving_2024}, we found that cluster stars typically have a dispersion in proper motion of $\sim$0.2--0.3~mas~yr$^{-1}$. In contrast, stars in OB associations in the catalogues of \citet{chemel_search_2022} and \citet{melnik_kinematics_2017} are less bound, showing a proper motion dispersion of $\sim$1--2~mas~yr$^{-1}$ (Fig.~\ref{fig:6cas_offset_and_pm}). Thus, depending on the stellar environments surrounding each target, the cut radius varies from 0.1 to 1.5~\masyr.

A YSG/YHG may still belong to a stellar group even if no nearby co-moving stars are identified in our search. We estimated the limiting distance at which hot stars would still fall within our criteria. 
As a representative case, we considered a B3 main-sequence star with $T_\mathrm{eff}=17\,000$~K and an absolute $G$-band magnitude $M_G=-1.19$~mag.\footnote{\url{http://www.pas.rochester.edu/~emamajek/EEM_dwarf_UBVIJHK_colors_Teff.txt}} We adopted an average visual extinction in the Galactic disc of $A_G \approx 1$~mag~kpc$^{-1}$. The extinction value varies strongly with direction and the local extinction law: in the solar neighbourhood $A_V\approx 1$~mag~kpc$^{-1}$ and $0.80 \leq A_G/A_V \leq 0.89$ for $2\leq R_V \leq 4$ \citep{gontcharov_new_2023}. The average extinction of open clusters in the Galactic disc is $A_V=0.70$~mag~kpc$^{-1}$ \citep{froebrich_old_2010},  but the extinction could be as low as $A_V=0.37$~mag~kpc$^{-1}$ in diffuse regions \citep{wang_opticalmid-infrared_2017} or higher in specific directions, e.g. $A_V=1.45$~mag~kpc$^{-1}$ for open cluster King~7 \citep{straizys_interstellar_2021}. 
We adopted an upper magnitude limit of $G=18$~mag, as fainter stars are unlikely to have reliable parallaxes. From the absolute magnitude equation $G=M_G+5\log d -5 +A_G$, where $d$ is distance in parsecs, we find that a B3 star has an apparent $G$-band magnitude of 9.8~mag at 1000~pc, 14.2~mag at 3000~pc, and 15.8~mag at 4000~pc. At the cooler end of our hot star selection, an A2 main-sequence star with $T_\mathrm{eff}=8\,800$~K and absolute magnitude $M_G=1.35$~mag would appear at $G$-band magnitudes of 12.3~mag at 1000~pc, 16.7~mag at 3000~pc, and 18.4~mag at 4000~pc. Because the hotter end of the temperature range contains fewer stars, this method is generally able to detect stellar populations out to distances of 3000--4000~pc.

\subsection{Example}\label{sect:dist_example}

\begin{figure}[]
    \centering
    \includegraphics[width=0.85\linewidth]{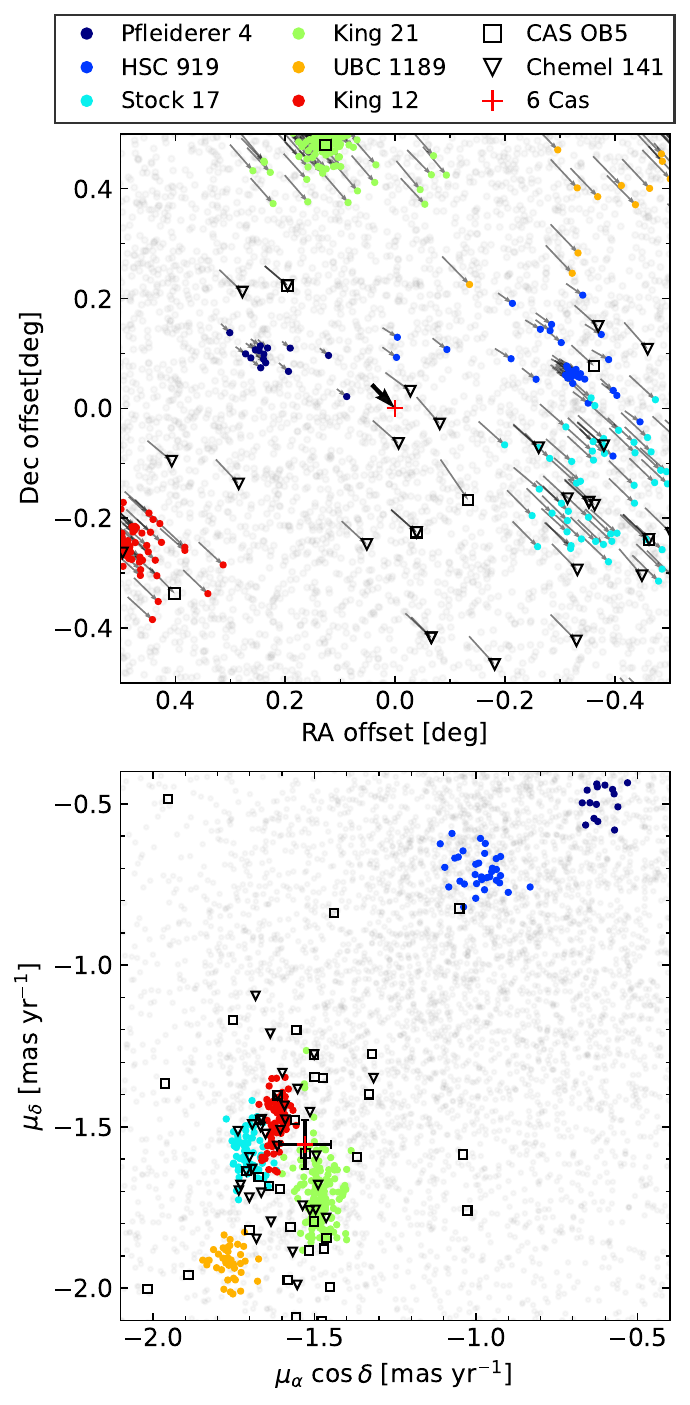}
    \caption{Sky region around 6\,Cas (red cross). Top panel: projected offsets on sky. Bottom panel: proper motions. Field stars from \textit{Gaia} are marked in grey (one in ten has been plotted), open clusters from \citet{hunt_improving_2024} are marked in coloured points, members of the OB associations Cas\,OB5 from \citet{melnik_kinematics_2017} and no.~141 \citep{chemel_search_2022} are marked with open symbols (some members overlap between the two catalogues). Arrows in the top panel indicate the motion of cluster and OB association stars over the past 0.1\,Myr based on their proper motions. }
    \label{fig:6cas_offset_and_pm}
\end{figure}

As an illustrative example, we applied this method to the YHG 6\,Cas. 
A previous study by \citet{maiz_apellaniz_lucky_2021} identified three nearby stars with reliable \textit{Gaia} DR2 astrometry and derived a distance of $2780^{+370}_{-290}$~pc by combining their parallaxes. 
Figure~\ref{fig:6cas_offset_and_pm} shows the sky region surrounding 6\,Cas, including the positions and proper motions of nearby stars, clusters, and OB associations. 6\,Cas has similar proper motion with three clusters, and OB associations Cas\,OB5 \citep{melnik_kinematics_2017} and no.~141 \citep{chemel_search_2022}. 

\begin{figure}[h]
    \centering
    \includegraphics[width=\linewidth]{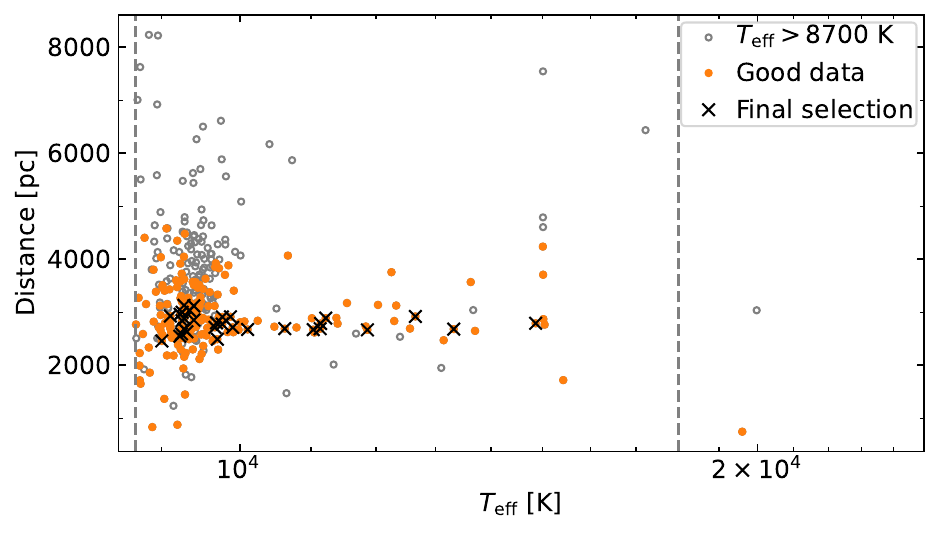}
    \caption{\textit{Gaia} effective temperatures and \citet{bailer-jones_estimating_2021} distances for hot stars within 10{\arcmin} of 6\,Cas. Stars meeting our astrometric quality criteria are highlighted in orange. The final selection of stars used for the distance estimation is marked with black crosses. }
    \label{fig:6cas_example_teff_dist}
\end{figure}

We selected stars from \textit{Gaia} DR3 within a 10{\arcmin} on-sky radius of  6\,Cas, following the criteria listed above. The temperature criterion is informative. Figure~\ref{fig:6cas_example_teff_dist} shows the distances from \citep{bailer-jones_estimating_2021} plotted against \textit{Gaia} effective temperatures for hot stars within 10{\arcmin} of 6\,Cas. Stars with lower temperatures have a large scatter in distance, while stars hotter than 10\,000~K seem to form a plateau just below 3000~pc. 

\begin{figure}
    \centering
    \includegraphics[width=0.8\linewidth]{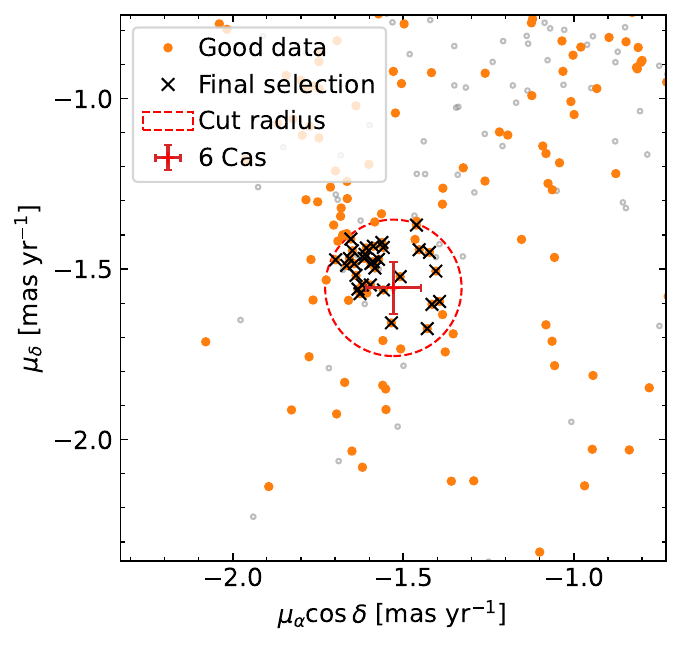}
    \caption{Selection of stars near 6\,Cas (red cross) based on proper motion, using the same sample as in Fig.~\ref{fig:6cas_example_teff_dist}. The proper motion cut with a radius of 0.2~mas~yr$^{-1}$ is marked with a red circle. }
    \label{fig:6Cas_pmcut}
\end{figure}

After correcting for the proper motion bias, we selected stars with proper motions similar to 6\,Cas, i.e., stars with proper motions within 0.2~mas~yr$^{-1}$ of its value (see Fig.~\ref{fig:6Cas_pmcut}). Applying the proper motion cut results in 38 stars. 

We discarded any obvious outliers and stars whose parallaxes deviate by more than $2\sigma$ ($\gtrsim 0.1$~mas) from the sample mean value, resulting in a final selection of 31 stars. These stars are marked with black crosses in Fig.~\ref{fig:6cas_example_teff_dist} and Fig.~\ref{fig:6Cas_pmcut}. Their consistent distances suggest that they belong to a common, association-like environment. A YSG/YHG that is no longer associated with its birth environment would not show such a tight distance grouping. 

We corrected the parallaxes of the final sample for the \textit{Gaia} zero point offset \citep{lindegren_gaia_2021-1}. These offsets are small ($\sim$0.02--0.04~mas). We combined the parallaxes following the recipe of \citet{maiz_apellaniz_validation_2021}. The resulting mean parallax for 6\,Cas is $\varpi = 0.3586 \pm 0.0088$. We derived a geometric distance for 6\,Cas of $2790_{-71}^{+70}$~pc, where the errors correspond to the 68\% confidence interval. This result agrees well with the previous estimate of $2780_{-290}^{+370}$~pc by \citet{maiz_apellaniz_lucky_2021}. 

Given the similarity in proper motion and spatial proximity to Cas~OB~5, located at a distance of 2500--3000~pc \citep{quintana_quantifying_2025}, we support the conclusion of \citet{melnik_kinematics_2017} that 6\,Cas is likely a member of this association. For the same reasons, 6\,Cas may belong to association no.~141 identified by \citet{chemel_search_2022} at a distance of 2827~pc.
The proper motion, on-sky location, and distance of Cas~OB5 members identified by \cite{melnik_kinematics_2017} largely overlap with stars in association no.~141 from \cite{chemel_search_2022}, suggesting that they are the same  stellar group (see Fig.~\ref{fig:6cas_offset_and_pm}).

\longtab{
\begin{longtable}{l|p{14mm}|p{14mm}p{15mm}p{5mm}|p{19mm}p{14mm}p{5mm}|p{14mm}p{16mm}}
\caption{\label{tab:distances} Distances and cluster/OB association memberships for YSGs/YHGs. The targets are ordered by increasing right ascension. 'Stellar group' indicates that the star is not associated with any known cluster or OB association, but a co-moving group of nearby stars is identified. }\\
\hline\hline
Identifier &  Distance \textit{Gaia}$^{(4)}$ (pc) & Lit. cluster/assoc. & Lit. distance~(pc) & Ref. & Ident. cluster/assoc. &  Distance~to cl/assoc. (pc) & Ref. & \ion{H}{I} kinematic distance (pc) & Group distance (pc) \\
\hline
\endfirsthead
\caption{continued.}\\
\hline\hline
Identifier &  Distance \textit{Gaia}$^{(1)}$ (pc) & Lit. cluster/assoc. & Lit. distance~(pc) & Ref. & Ident. cluster/assoc. &  Distance~to cl/assoc. (pc) & Ref. & \ion{H}{I} kinematic distance (pc) & Group distance (pc) \\
\hline
\endhead
\hline
\endfoot
$\phi$\,Cas       &$3978_{-903}^{+1627}$ & NGC\,457 & 2420 & 5,6 & NGC\,457 &$2800\pm9$&2& \mbox{1000-1800} & $2920^{+114}_{-113}$ \\
HD\,10494         &$2823_{-114}^{+142}$  & NGC\,654 \newline Cas OB8 & 2450 \newline  2300 & 5,6 \newline 7 & NGC\,654 \newline Chemel\,150 &$2815\pm11$ \newline 2626 &2\newline 3 & \mbox{1000-1800} & $2784_{-61}^{+57}$\\
HD\,12399         &$3792_{-192}^{+240}$  & - & - && - & - && \mbox{2000-3200} & $3564_{-266}^{+320}$\\
HD\,18391         &$2263_{-93}^{+113}$  & Anon. Cluster & $1661\pm73$ & 8 & SAI\,25 \newline Chemel\,143 & $2252\pm11$ \newline 2262 &2\newline 3 & \mbox{2000-2800} & $2242_{-52}^{+49}$ \\
$\epsilon$ Aur    &$1056_{-182}^{+218}$  & Aur OB1 \newline - \newline - & 1060 \newline 900 \newline $1500\pm500$ & 5,7 \newline 9 \newline 10 & Chemel\,147 & 1154 & 3 & $\sim$700 & $1127\pm31$ \\
HD\,57118         &$2759_{-138}^{+116}$  & - & - && Chemel\,25 & 2524 & 3 & $\sim$3300 & $2790_{-93}^{88}$ \\
R\,Pup            &$3958_{-228}^{+291}$  & NGC\,2439 & 4450 & 5,11 & NGC\,2439 \newline Chemel\,68 & $3350\pm13$ \newline 3237 &2\newline 3 & $\sim$3500 & $3331_{-80}^{+77}$  \\
HD\,74180         &$2532_{-397}^{+600}$& Vel\,OB1 \newline Vel\,OB1 & 1460 \newline \mbox{$1750\pm156$} & 7,12 \newline 13 & Chemel\,107 & 1935 & 3 & \mbox{1500-2000} & $1657_{-34}^{+33}$ \\
HD\,75276         &$1443_{-57}^{+74}$    & Vel\,OB1 \newline Vel\,OB1 & 1460 \newline \mbox{$1750\pm156$} & 7 \newline 13 & Chemel\,106 & 1447 & 3 & $\sim$1500 & $1538\pm24$\\
V709\,Car         &$4006_{-376}^{+397}$  & - & - && Stellar group &&& $\sim$3600 & $3817_{-271}^{+315}$\\
TYC\,8958-479-1   &$8783_{-1330}^{+2122}$& Barbá\,2 & $7390^{+650}_{-550}$ & 14 & Chemel\,8 & 6917 & 3 & - & $7025_{-572}^{+687}$\\
HD\,96918         &$4583_{-1127}^{+1376}$& Car\,OB2 \newline Car\,OB2 \newline - & 1790 \newline 2000 \newline \mbox{$2700\pm1000$} & 7 \newline 12 \newline 15 & Chemel\,210/211 & 2362/2450 & 3 & \mbox{500/5000} & $2623_{-83}^{+88}$ \\
{$o^1$}\,Cen      &$3783_{-928}^{+3197}$  & - & - && - & - && $\sim$2800 & $2880\pm100$\\
V810\,Cen         &$5312_{-1319}^{+2095}$& Cru\,OB1 \newline Stock\,14 & 2010 \newline $2780\pm120$ & 7 \newline 16 & Chemel\,205 \newline Stock\,14 & 2347 \newline $2340\pm9$ & 3 \newline 2 & \mbox{2400-3500} & $2330_{-65}^{+61}$\\
HR\,5171A         &$3601_{-539}^{+649}$  & Gum 48d \newline R 80 \newline R 80 \newline - & 3500 \newline 3600 \newline 2900 \mbox{$1500\pm500$} & 17 \newline 12 \newline 7 \newline 18 & SFR & - && $\sim$3500 & $2953_{-96}^{+92}$$^{(a)}$ \\
UCAC2\,4867478    &$3327_{-331}^{+465}$  & - & - && Stellar group & - && $\sim$3600 & $3322_{-314}^{+351}$\\
IRAS\,14394-6059  &$5450_{-686}^{+770}$ & - & - && - & - &  & - & -* \\
CD-59\,5594       &$3739_{-279}^{+325}$  & - & - && Pismis\,21 \newline Chemel\,47 & $2906\pm24$ \newline 2958 &2\newline 3 & $\sim$2500 & $2849\pm77$ \\
HD\,144812        &$1352_{-58}^{+66}$    & - & - && -  & - && - & $1374_{-132}^{+169}$$^{(a)}$  \\
V870\,Sco         &$3945_{-642}^{+940}$  & NGC\,6231 backgr. &  $<2090$ & 19 & - & - && $\sim$3500 & -$\dagger$ \\
V915\,Sco         &$1717_{-115}^{+175}$  & - \newline - & $2630^{+390}_{-339}$ \newline \mbox{$2800\pm 1100$} & 20 \newline 21 & HSC\,2870 \newline Chemel\,82 & $2350\pm25$ \newline 2250 &2\newline 3 & \mbox{1500-2000} & $2253_{-56}^{+57}$  \\
IRAS\,17163-3907  &$5196_{-1032}^{+1401}$ & - \newline - \newline RCW 122 \newline - & $1200^{+400}_{-200}$\newline \mbox{3600-4700} \newline $3380\pm300$ \newline $\sim$1000 & 22 \newline 23 \newline 24 \newline 24 & - & - && $\sim$1000 & -$\dagger$ \\
V925\,Sco         &$3006_{-166}^{+192}$  & Trumpler\,27 \newline Trumpler\,27 & 2427 \newline $2100\pm200$ & 25 \newline 26 & Chemel\,82 & 2250 & 3 & \mbox{1800-4000} & $2310_{-110}^{+126}$\\ 
{[FMR2006]}\,15   &$3654_{-1750}^{+1743}$ & RSGC\,1 \newline RSGC\,1 & 5800 \newline 6600 & 27 \newline 28 & RSG group & - && $\sim$6500 & -$\dagger$ \\
IRAS\,18357-0604  &$4539_{-1655}^{+1939}$  & RSG\,assoc. & $\sim$6000 & 29 & RSG assoc. & - && $\sim$5500 & -$\dagger$ \\
HD\,179821        &$4432_{-355}^{+349}$  & - \newline - \newline - & 6000 \newline \mbox{$6000\pm1000$} \newline $\geq4000$ & 18 \newline 30 \newline 31 & - & - && >5000 & -*\\
V1452\,Aql        &$2748_{-148}^{+193}$  & - & - && CWNU\,1591 & 2290 & 4 & \mbox{2500-3500} & $2472_{-104}^{+111}$\\
IRC\,+10420       &$4260_{-752}^{+878}$ & - \newline - & 5800 \newline \mbox{4000-6000} & 32 \newline 33 & - & - && $\sim$5500 & -* \\
V1027\,Cyg        &$3723_{-237}^{+265}$  & - & - && Chemel\,15 & 3902 & 3 & \mbox{2000-4500} & $3977_{-208}^{+226}$\\
HD\,331777        &$4559_{-363}^{+388}$  & - & - && Kronberger\,54 \newline Chemel\,15 & $4025^{+73}_{-70}$ \newline 3902 &2\newline 3 & \mbox{2000-5000} & $3789_{-176}^{+190}$\\
RW\,Cep           &$6666_{-1006}^{+1561}$ & Berkeley\,94 \newline Cep\,OB1 & $3900\pm110$ \newline 3470 & 34 \newline 12 & Berkeley\,94 \newline Chemel\,122 & $4000\pm40$ \newline $3876$ & 2 \newline 3 & $\sim$3500 & $3921_{-157}^{+168}$\\
V509\,Cas         &$3917_{-737}^{+969}$  & Cep\,OB1 \newline Cep\,OB1 \newline - & 2780 \newline 3470 \newline \mbox{$1370\pm480$} & 7 \newline 12 \newline 35 &  Chemel\,120 & 3055 & 3 & $\sim$3800 & $3368\pm127$\\
6\,Cas            &$2319_{-288}^{+463}$  & Cas\,OB5 \newline -  & \mbox{2500-3000} \newline $2780^{+370}_{-290}$ & 7,36 \newline 37 &  Chemel\,141 & 2827 & 3 & \mbox{2000-3000} & $2790_{-71}^{+70}$ \\
HD\,223767        &$2879_{-117}^{+136}$  & Cas\,OB5 & \mbox{2500-3000} & 7,36 &  Chemel\,141 & 2827 & 3 & \mbox{2000-3000} & $2740_{-83}^{+86}$ \\
$\rho$\,Cas       &$6747_{-1647}^{+1918}$& Cas\,OB5 \newline - \newline - & \mbox{2500-3000} \newline \mbox{$2500\pm500$} \newline $3100\pm500$ & 7,36 \newline 18 \newline 38 &- &-&& \mbox{2500-3200} & $2810_{-102}^{+104}$ \\
\end{longtable} 
\tablefoot{
* Use \cite{bailer-jones_estimating_2021} distance. 
$\dagger$ Use \ion{H}{I}-based distance. 
\tablefoottext{a}{See comments in App.~\ref{sect:distance_comments_stars}.}
}
\tablebib{(1)~\citet{bailer-jones_estimating_2021}; (2)~\citet{hunt_improving_2024}; (3)~\citet{chemel_search_2022}; (4)~\citet{he_unveiling_2023}; (5)~\citet{arellano_ferro_colour_1990}; (6)~\citet{rastorguev_statistical_1999}; (7)~\citet{melnik_kinematics_2017}; (8)~\citet{turner_cepheid_2009}; (9)~\citet{strassmeier_time-series_2014}; (10)~\citet{guinan_large_2012}; (11)~\citet{white_ubv_1975}; (12)~\citet{humphreys_studies_1978}; (13)~\citet{reed_vela_2000}; (14)~\citet{maiz_apellaniz_barba_2025}; (15)~\citet{achmad_photometric_1992}; (16)~\citet{turner_new_1982}; (17)~\citet{karr_gum_2009}; (18)~\citet{van_genderen_pulsations_2019}; (19)~\citet{damiani_chandra_2016}; (20)~\citet{andrews_hr_1977}; (21)~\citet{vasquez_h_2005}; (22)~\citet{koumpia_optical_2020}; (23)~\citet{lagadec_double_2011}; (24)~\citet{wallstrom_investigating_2015}; (25)~\citet{perren_photometric_2012}; (26)~\citet{moffat_trumpler_1977}; (27)~\citet{figer_discovery_2006}; (28)~\citet{davies_cool_2008}; (29)~\citet{clark_iras_2014}; (30)~\citet{hawkins_discovery_1995}; (31)~\citet{reddy_spectroscopic_1999}; (32)~\citet{nedoluha_vla_1992}; (33)~\citet{jones_irc_1993}; (34)~\citet{delgado_berkeley_2013};  (35)~\citet{nieuwenhuijzen_hypergiant_2012}; (36)~\citet{quintana_quantifying_2025}; (37)~\citet{maiz_apellaniz_lucky_2021}; (38)~\cite{zsoldos_photometry_1991}
}
}

\subsection{Results}

\begin{figure}
    \centering
    \includegraphics[width=\linewidth]{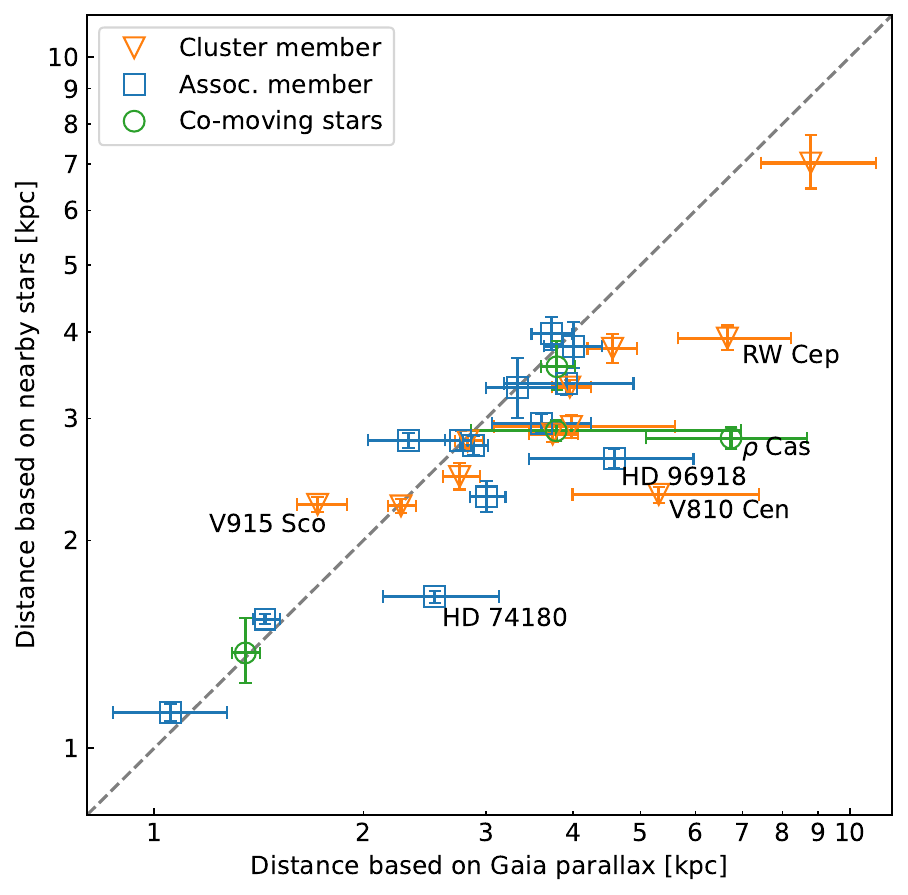}
    \caption{Comparison of group-based distances derived in this work with distances derived from \textit{Gaia} parallaxes by \cite{bailer-jones_estimating_2021}. Stars associated with clusters, OB association, and stars without an origin group are marked with different symbols and colours. The grey dashed line marks the one-to-one relation. Outliers are marked. Both axes are logarithmic. }
    \label{fig:distance_comparison}
\end{figure}

The group-based distances derived in this work are listed in Table~\ref{tab:distances}. The proper motion cuts are illustrated in App.~\ref{app:prop_mot_fig} and the stars used for combining the parallaxes are listed in App.~\ref{apptable:nearby_stars_info} with a full table available electronically. 

Figure~\ref{fig:distance_comparison} compares our group-based distances for the YHG/YSG sample with those listed in the \cite{bailer-jones_estimating_2021} catalogue. We used the \textsc{rgeo} values from that catalogue, as the \textsc{rpgeo} distances that consider the photometric colour indices are unlikely to reliably characterise Supergiants. 
For stars with relatively small parallax errors, our results agree well with the \cite{bailer-jones_estimating_2021} catalogue values. We found significant differences for stars that have very large parallax errors, where the \cite{bailer-jones_estimating_2021} distances tend to be very large, above 4000~pc. We found that they are likely much closer, because we identified possible affiliations with clusters or OB associations at distances of 2000--3000~pc.

Overall, we improved the distance estimates for 28 stars out of the 35 stars in our sample. 
Of these 28 stars, 11 are cluster members, 13 are in OB associations, and four stars do not belong to a stellar group, but we found a co-moving population. For the remaining seven stars, we were unable to determine a group-based distance. IRAS\,17163-3907, HD\,179821, IRC\,+10420, V870\,Sco, and IRAS\,14394-6059 have co-moving stars spanning distances from 1000~pc to 4000~pc, but no clearly identifiable origin group. Most of these stars are also too distant or too reddened for this method. The final two stars, [FMR2006]\,15 and IRAS\,18357-0604, belong to a distant ($\sim$6000~pc) stellar population rich in RSGs. We were unable to obtain group-based distances for them because of their large distance and high extinction.

\section{Kinematic distances from the Galactic \ion{H}{I} map}\label{sect:hi_dist}

To independently verify the group-based distances derived in Sect.~\ref{sect:distance}, we compared the stellar radial velocities of our targets with the kinematics of \ion{H}{I} gas in the Milky Way. The \ion{H}{I} gas traces both the small-scale and large-scale structures in the Galaxy \citep[][and references therein]{mcclure-griffiths_atomic_2023}. If a star is affiliated with a cluster or an OB association, which are generally moving together with the Galactic rotation and the surrounding interstellar medium \citep[e.g.][]{castro-ginard_milky_2021}, its group-based distance should be consistent with the distance inferred from the kinematics of the surrounding \ion{H}{I} gas. The radial velocities of Supergiants and stars younger than $10^8$~yr are correlated with the \ion{H}{I} gas velocity \citep{fletcher_model-independent_1963,humphreys_space_1970}. 

\subsection{Method}

We used the spatially coherent 3D kinematic map of the \ion{H}{I} gas in the Milky Way by \cite{soding_spatially_2025}, based on 21-cm emission measurements from the HI4PI survey \citep{bekhti_hi4pi_2016}. The map is in Hierarchical Equal Area isoLatitude Pixelation of a sphere (HEALPix) format. It includes eight posterior samples for each point in the Sun-centred HEALPix-times-distance grid. We used the auxiliary fields data product by \cite{soding_spatially_2024}, which includes the three components of the Galactic velocity field in each grid point. The velocity field is heliocentric, with the peculiar motion of the Sun corrected for. The bin size increases from $\sim$30~pc at 2000~pc distance to $\sim$100~pc at 5000~pc distance.

For each star in our sample, we extracted the line-of-sight radial velocity of the \ion{H}{I} gas within a $\sim$0.5~deg radius and plotted its value as a function of distance. We compared these gas velocity profiles with the stellar systemic radial velocity to determine the distance where both values are equal. 
We used systemic radial velocities from monitoring studies or mean radial velocities and variability amplitudes from \textit{Gaia} DR3, where the former were not available. The adopted stellar radial velocities are listed in Table~\ref{app:lit_rvs}. 
The radial-velocity variability of YHGs makes it difficult to determine accurate systemic velocities, and the gas velocity map provides accuracy at scales of hundreds of parsecs \citep{soding_spatially_2025}. Therefore, the resulting distances have uncertainties of the order of $\sim$500~pc. 

\subsection{Examples}

\begin{figure}
    \centering
    \includegraphics[width=\linewidth]{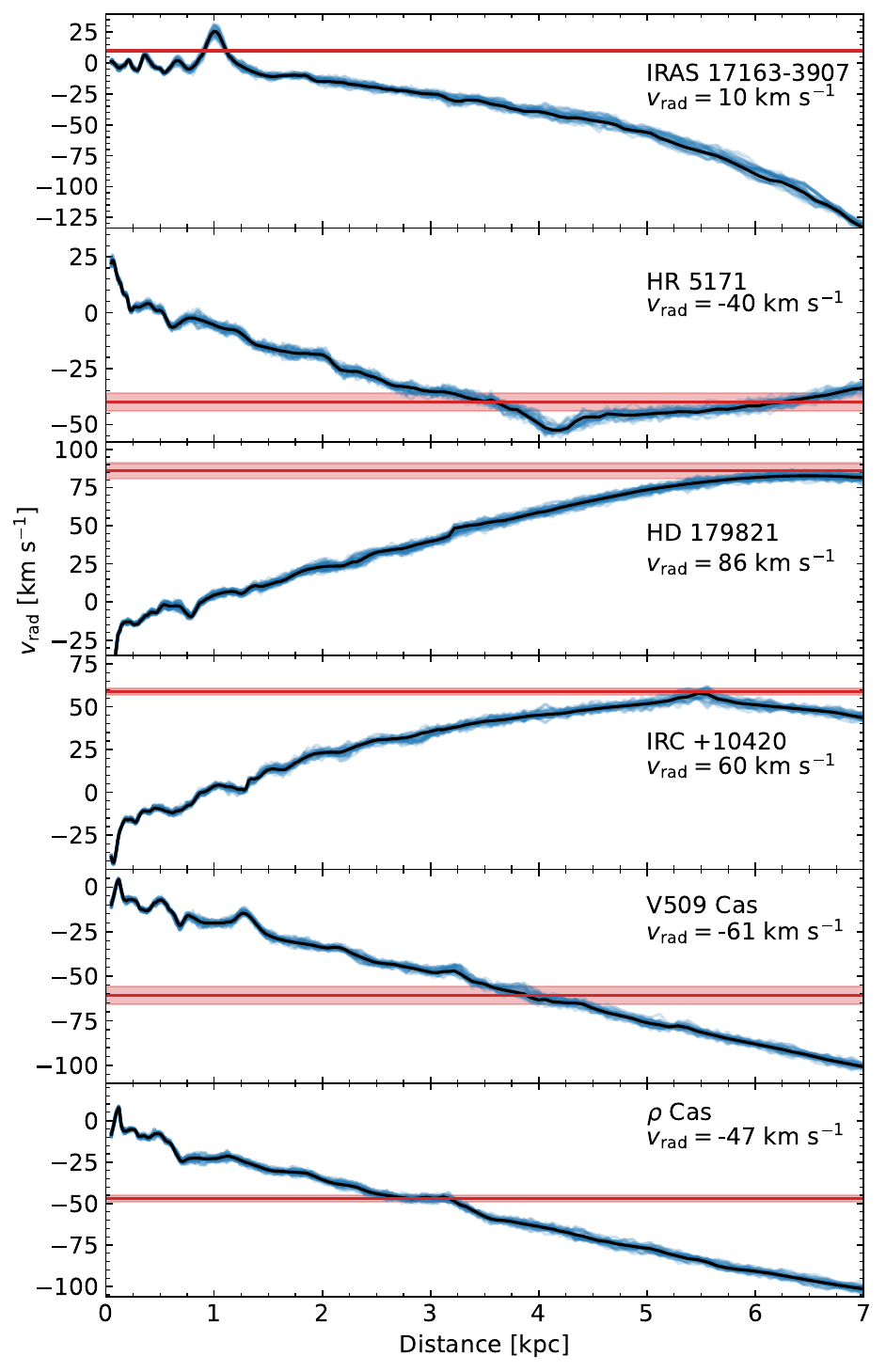}
    \caption{Radial velocity of \ion{H}{I} gas along the lines of sight towards six YHGs. The thin blue lines show the sampled posterior velocity distribution of  \ion{H}{I} at each distance bin within a 0.5~deg region around each star. The solid black line indicates the mean velocity. Red horizontal lines mark the observed stellar radial velocity, with the shaded region showing the approximate variability amplitude.  }
    \label{fig:HIyhgs}
\end{figure}

The gas velocity profiles and $v_\mathrm{rad}$ for six well-known YHGs are shown in Fig.~\ref{fig:HIyhgs}. Depending on the shape of the \ion{H}{I} velocity curve in the line-of-sight towards each star, we obtained either a single distance estimate when the stellar systemic velocity equals the gas velocity at a specific distance (e.g., V509\,Cas), or a distance range when the shape of the \ion{H}{I} velocity curve is flat over a distance span (e.g., $\rho$\,Cas). We highlight a few YHGs that are located in different Galactic lines-of-sights.

\textit{IRAS~17163-3907}. The radial velocity of IRAS~17163-3907 with respect to the Local Standard of Rest (LSR) is 18~\kms \citep{wallstrom_investigating_2015}. Transformation to the heliocentric reference frame results in a radial velocity of 9.7~\kms, which is compatible with the \ion{H}{I} gas velocity near 1000~pc. We were unable to constrain a group-based distance. The \ion{H}{I}-based distance agrees well with the $1200^{+400}_{-200}$~pc distance estimate from the Gaia DR2 parallax \citep{koumpia_optical_2020} and from spectrophotometric analysis \citep{wallstrom_investigating_2015}. It differs significantly from the 3600--4700~pc range proposed by \cite{lagadec_double_2011}, and from the distance derived from the \textit{Gaia} DR3 parallax ($5196^{+1401}_{-1032}$~pc; \citealt{bailer-jones_estimating_2021}). Our result supports a shorter distance of $\sim$1000~pc for IRAS~17163-3907.

\textit{HR 5171}. The radial velocity of HR\,5171 has been extensively monitored, with mean values of $-40$~\kms\ \citep{humphreys_high-luminosity_1971} and $-38$~\kms\ with a variability amplitude of 3.5~\kms\ \citep{balona_radial_1982}. This velocity corresponds to \ion{H}{I} gas at a distance of 3500~pc, similar to the group-based distance of $2953^{+92}_{-96}$~pc within errors. The \ion{H}{I}-based distance is also consistent with the established distance of 3600~pc \citep{humphreys_high-luminosity_1971,chesneau_yellow_2014} and with the distance from the \textit{Gaia} parallax of $3601^{+649}_{-539}$~pc \citep{bailer-jones_estimating_2021}, and does not support the $1500\pm500$~pc distance suggested by \cite{van_genderen_pulsations_2019}. For HR\,5171, a more distant distance solution at $\sim$6000~pc is also possible. Such a distance ambiguity is possible for objects located in the inner Galaxy inside the solar orbit \citep{nakanishi_three-dimensional_2003}. In these cases, the alternative solution is generally very small (<1000~pc) or very large (>5000~pc) and unlikely for most targets. We do not list them in Table~\ref{tab:distances}. Our results confirm the distance of HR\,5171 at 2900--3500~pc. 

\textit{HD 179821}. The high radial velocity of HD\,179821 at 86~\kms\ \citep{sahin_hd_2016} shows good alignment with the gas velocity only at large distances (>5000~pc), providing a lower distance limit. We were unable to constrain a group-based distance. Literature values vary from 1000~pc \citep{josselin_probing_2001} to 6000~pc \citep{van_genderen_pulsations_2019}. The \textit{Gaia} DR3 parallax is reliable, with an error of $\sim$10\% and a \textsc{ruwe} value of 0.92, corresponding to a distance of $4432_{-355}^{+349}$~pc \citep{bailer-jones_estimating_2021}, which is in acceptable agreement with our result. We adopted the \textit{Gaia} distance for this star. 

\textit{IRC +10420}. The radial velocity measured from emission lines in IRC\,+10420 is between 60 and 68~{\kms} \citep{klochkova_optical_1997,humphreys_crossing_2002}, consistent with the result of CO rotational lines of $v_\mathrm{LSR}~ 75$~{\kms} (LSR correction $\sim$16~\kms; \citealt{oudmaijer_high_1998}). This velocity suggests a very large distance of 5500~pc. We were unable to constrain a group-based distance. The high distance is consistent with earlier estimates of 5800~pc \citep{nedoluha_vla_1992} and 4000--6000~pc  \citep{jones_irc_1993} derived from Galactic rotation models. Our result is $\sim$1000~pc greater than the distance derived from the \textit{Gaia} parallax ($4260^{+878}_{-752}$~pc). We adopted the \textit{Gaia} distance for this star. 

\textit{V509\,Cas}. The star has a systemic velocity of $-60.7$~\kms\ \citep{kasikov_yellow_2024}, corresponding to a \ion{H}{I}-based distance of $\sim$3800~pc. This agrees with the group-based distance of $3368\pm127$~pc within the errors. This result is significantly different from $1370\pm480$~pc based on the parallax measured by the \textit{Hipparcos} satellite and commonly used for this star \citep{nieuwenhuijzen_hypergiant_2012,van_genderen_pulsations_2019}, but in good agreement with a distance of $3917^{+969}_{-737}$~pc based on its \textit{Gaia} parallax  \citep{bailer-jones_estimating_2021}. Thus, we adopted the group-based distance. 

\textit{$\rho$ Cas}. The long-term radial velocity monitoring of $\rho$\,Cas gives a well-constrained systemic velocity of $-47\pm2$~\kms\ \citep{lobel_high-resolution_2003}. This matches the \ion{H}{I} velocity at a distance between 2500~pc and 3200~pc, in good agreement with our estimated group-based distance of $2810_{-102}^{+104}$~pc and the $2500\pm500$~pc distance from \cite{van_genderen_pulsations_2019}. Thus, we adopted the group-based distance.

\subsection{Results and comparison with group-based distances}\label{sect:discussion_distance}

The \ion{H}{I}-based distance results are listed in Table~\ref{tab:distances}. We derived \ion{H}{I}-based distances for 32 stars, six of them lack a group-based distance estimate due to high distance and extinction. For three stars systemic radial velocities were not available. One target, IRAS\,14394-6059, lacks both types of distance estimates, but the \textit{Gaia} parallax places the star beyond 5000~pc, making it a potentially luminous YSG. 

Six stars have only \ion{H}{I}-based distances. Three of them, IRC\,+10420, HD\,179821, and IRAS~17163-3907, were included in the examples above. The other three stars are [FMR2006]\,15, IRAS 18357-0604, and V870\,Sco. In general, the \ion{H}{I}-based distances agree with the distances derived from the Galactic rotation. Information for all stars is included in App.~\ref{sect:distance_comments_stars}. 

\begin{figure}[]
    \centering
    \includegraphics[width=\linewidth]{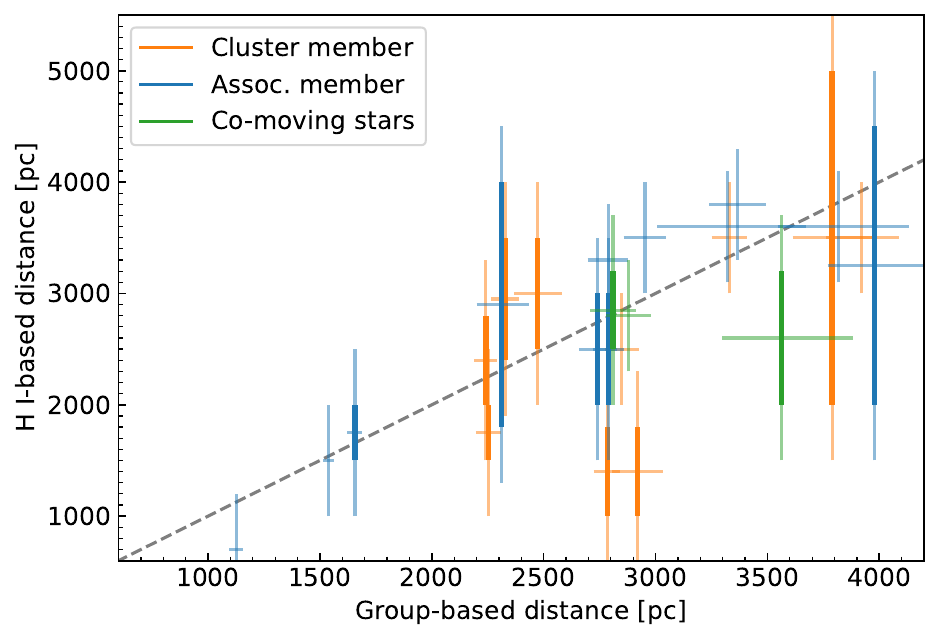}
    \caption{Comparison of group-based and \ion{H}{I}-based distances. Solid lines indicate the \ion{H}{I}-based distance ranges and lighter lines illustrate the uncertainties. Affiliations with stellar groups are indicated with different colours as in Fig.~\ref{fig:distance_comparison}. The grey dashed line marks the one-to-one relation. }
    \label{fig:group_vs_hi_dist}
\end{figure}

Generally, the results of \ion{H}{I}-based distances coincide well with the group-based distances within the uncertainties of about 500~pc (see Fig.~\ref{fig:group_vs_hi_dist}). For the group-based distances the uncertainties are of the order of 100~pc for most stars. Group-based distances are most reliable when a proper motion alignment with a cluster or OB association can be established. Otherwise, there remains a higher probability that stars with similar proper motions do not represent the birth environment. Coincidental proper motion alignment between a YSG/YHG and an unrelated OB association or open cluster at a different distance is possible. This problem becomes more severe toward the Galactic centre, where multiple stellar populations lie along the same line of sight, complicating the identification of a possible origin population and preventing reliable distance determination for several stars. Differences between the \ion{H}{I}-based and group-based distances can indicate an incorrect origin group association or help identify potential runaway stars. Of the 26 stars that have both distance estimates, 23 stars have good agreement between the results. For three stars ($\phi$\,Cas, HD\,10494, HD\,96918), we find significant discrepancies detailed below.

\textit{$\phi$\,Cas and HD\,10494}. The two stars have similar radial velocities and are within $5~\deg$ distance from each other on the sky, which corresponds to a physical separation of 270~pc using their group-based distances. The group-based distances of both stars are close to 2800~pc and are based on alignment with young clusters. For HD\,10494, the group-based distance is also in good agreement with the distance derived from the \textit{Gaia} DR3 parallax, while the parallax of $\phi$\,Cas is unreliable. The radial velocities of $\phi$\,Cas ($-28$~{\kms}) and HD\,10494 (around $-30$~{\kms}) are very similar to their home clusters NGC\,457 and NGC\,654 with $v_\mathrm{rad}\approx -34$~\kms \citep{rastorguev_statistical_1999}. The radial velocities differ from the \ion{H}{I} velocity at 2800~pc by 15--20~\kms and correspond to \ion{H}{I}-based distances in the range of 1000--1800~pc. This discrepancy between \ion{H}{I}-based and group-based distances could be a local phenomenon. These clusters are located in the Perseus arm of the Milky Way, in a region with many other young clusters. They are potential members of the Cassiopeia-Perseus open cluster complex, sharing a common origin in the same giant molecular cloud \citep{de_la_fuente_marcos_cassiopeia_2009}. If this is the case, the shocks from supernovae from the previous generation of stars could have affected the kinematics of stellar populations in the region. 

\textit{HD\,96918}. Comparison of its radial velocity with the \ion{H}{I} velocity map implies a distance of $\sim$500~pc or of $\sim$5000~pc. The field around HD\,96918 contains many stellar groups spanning from 500~pc to 4000~pc and aligning it with the kinematically closest OB association results in a distance of 2600~pc. \cite{achmad_photometric_1992} found the same distance ambiguity: Galactic rotation curve (500~pc or 5300~pc), interstellar reddening in line of sight ($2400\pm900$~pc), and comparing derived atmospheric parameters to evolutionary models (2200~pc). Our group-based distance result is in agreement with the latter two of their estimates. However, this discrepancy raises the possibility that the star is a runaway, which we discuss further in Section~\ref{sect:runaway}.

In the following sections, we used group-based distances when available. Otherwise, if \textit{Gaia} $\varpi/\sigma_\varpi>5$, we adopted the distances from \cite{bailer-jones_estimating_2021}. When neither of these are reliable, we used the \ion{H}{I}-based distance. 

\section{Radius and luminosity}\label{sect:rad_lum}

Based on our distance estimates, we determined the radii and luminosities for 15 YSGs and five YHGs in our sample. In this analysis, we only included stars with available homogeneous angular diameter measurements and adopted literature {\teff} values.

For stars with multiple published {\teff} values, we adopted an average temperature, and for stars without spectroscopic temperature estimates, we adopted the temperature closest to their spectral type classification. Due to the inhomogeneous data and discrepancies resulting from different methods, we estimated typical temperature uncertainties of at least 200~K. For stars with multiple published \teff\ measurements, we adopted uncertainties to encompass the range of published values.
The effective temperatures of YHGs can be highly variable. In their quiescent state, they exhibit quasi-periodic pulsations with cycles of a few hundred days, with {\teff} variability of the order of a couple hundred degrees \citep[e.g.][]{kasikov_yellow_2024,van_genderen_investigation_2025}. Temperature changes of 1000--3000~K have been observed during outbursts and on longer timescales \citep{lobel_high-resolution_2003,de_jager_obstacle_1997,klochkova_optical_1997}. Several YHGs, such as $\rho$~Cas, V509\,Cas, and IRC\,+10420, have been photometrically stable for the past decade or more and their temperatures have been determined during these quiescent phases \citep{van_genderen_investigation_2025,nieuwenhuijzen_hypergiant_2012,koumpia_tracing_2022}. For the comparatively ``normal'' YSGs, the effective temperatures can differ by several hundred degrees, depending on the methods used. 

\begin{table*}
\caption[]{\label{tab:radii_lum} Literature data and derived radii and luminosities for the sample stars. Literature values are listed in columns labelled "Lit." together with their respective references.}
\centering
\begin{tabular}{l|p{15mm}p{4mm}p{16mm}p{4mm}|p{16mm}p{16mm}p{4mm}|p{24mm}p{24mm}}
\hline \hline
Identifier &  \raggedright Lit. ang. diam.~(mas) & Ref. & \mbox{Lit.  $T_\mathrm{eff}$} (K) & Ref. & Lit. radius $(R_\odot)$ & \raggedright Lit. luminosity $(\log L/L_\odot)$ & Ref. & Radius $(R_\odot)$ & Luminosity $(\log L/L_\odot)$ \\
\hline
$\phi$\,Cas       & $0.88\pm0.09$ & 1 & $7200\pm100$ \newline $7300\pm300$ \newline $7341\pm40$ & 2 \newline 3 \newline 4 & $263\pm34$ \newline $245\pm45$ & 5.23 \newline - & 2 \newline 3 & $276\pm29$ & $5.25_{-0.13}^{+0.11}$ \\
HD\,10494          & $0.63\pm0.05$ & 1 & $6672\pm40$ \newline 7127 & 4 \newline 5 &&&& $187\pm15$ & $4.83_{-0.12}^{+0.09}$\\
$\epsilon$ Aur    & $1.95\pm0.20$ & 1 & $7395\pm70$ & 6 & 300 & 5.37 & 6 & $236\pm25$  & $5.13_{-0.12}^{+0.09}$\\
HD\,57118         & $0.55\pm0.05$ & 1 & $7427\pm50$ & 7 &&&& $163\pm15$ & $4.86_{-0.10}^{+0.09}$\\
R\,Pup            & $0.68\pm0.06$ & 1 & $4100\pm68$ \newline 6500 & 8 \newline 9 &&&& $245\pm22$ & $4.98_{-0.1}^{+0.09}$\\
HD\,74180         & $1.82\pm0.15$ & 1 & 7240 \newline 7839 & 10 \newline 5 & - & 5.46 & 11 & $325\pm28$ & $5.5_{-0.11}^{+0.09}$\\
HD\,75276         & $0.70\pm0.05$ & 1 & 7100 \newline $6920\pm40$ & 5 \newline 8 &&&& $116\pm9$ & $4.38_{-0.09}^{+0.07}$\\
HD\,96918         & $2.18\pm0.23$ & 1 & $5625\pm312$ \newline $5200\pm200$ \newline 5729 \newline $5866\pm47$ & 12 \newline 13 \newline 5 \newline 8 & 485$^{(a)}$ \newline $700\pm250$ \newline - & \mbox{$5.33\pm0.03$} \newline $5.5\pm0.4$ \newline 5.58 & 12 \newline 13 \newline 11 & $616\pm69$ & $5.5_{-0.2}^{+0.13}$\\
{$o^1$}\,Cen      & $1.30\pm0.12$ & 1 & F7 Ia \newline F8 Ia0 & 14 \newline 15 &&&& $403\pm41$ & $5.33_{-0.21}^{+0.15}$\\
V810\,Cen         & $0.89\pm0.07$ & 1 & $5970\pm100$ & 16 & 420 & 5.3 & 16 & $222\pm20$ & $4.74_{-0.11}^{+0.09}$\\
HD\,144812        & $0.44\pm0.01$ & 1 & $6400\pm100$ & 17 &&&& $65\pm8$ & $3.81_{-0.15}^{+0.10}$\\
V925\,Sco         & $0.97\pm0.09$ & 1 & 7500 & 18 &&&& $242\pm25$ & $5.22_{-0.11}^{+0.09}$\\
HD\,179821        & $0.56\pm0.01$ & 1 & $7350\pm200$ \newline \mbox{4900-6760} \newline \mbox{5900-6800} \newline 6750 & 19 \newline 20 \newline 21 \newline 22 & 350$^{(b)}$ \newline 400--450  \newline $700\pm260$ \newline - \newline - \newline - & 5.5 \newline 5.1-5.2 \newline $5.49_{-0.19}^{+0.13}$ \newline $5.26_{-0.08}^{+0.06}$ \newline 5.47 \newline $5.5\pm0.4$ & 19 \newline 23 \newline 24 \newline 25 \newline 26 \newline 22 & $264\pm22$ & $5.13_{-0.18}^{+0.12}$\\
V1452\,Aql        & $0.69\pm0.06$ & 1 & $7032\pm50$ \newline 8035 & 7 \newline 5 &&&& $182\pm18$ & $4.98_{-0.15}^{+0.11}$\\
V1027\,Cyg        & $0.94\pm0.07$ & 1 & $3930\pm170$ \newline 5000 & 27 \newline 28 & 423 & 4.60 & 27 & $403\pm37$ & $4.78_{-0.29}^{+0.17}$ \\
HD\,331777        & $0.68\pm0.04$ & 1 & F8Ia & 29 &&&& $278\pm23$ & $4.96_{-0.11}^{+0.08}$\\
RW\,Cep           & $2.14\pm0.17$ \newline $2.44\pm0.02$ & 1 \newline 30 & 4200 & 30 & \mbox{900--1760} \newline - & - \newline 5.74 & 30 \newline 11 & $902\pm82$ & $5.36_{-0.13}^{+0.10}$\\
V509\,Cas         & $1.57\pm0.12$ \newline $1.24\pm0.03$ & 1 \newline 31 & \mbox{7500-8000} \newline $7800\pm200$ \newline $7900\pm200$ & 32 \newline 33 \newline 34 & 400--800 \newline 340 \newline - \newline -  & 5.2-5.6 \newline - \newline 5.58-5.7 \newline 5.43 & 32 \newline 35 \newline 11 \newline 26 & $567\pm111$ (JDSC) \newline $449\pm20$ (interf.) & $6.03_{-0.23}^{+0.15}$ (JDSC) \newline $5.83^{+0.05}_{-0.06}$ (interf.)\\
HD\,223767        & $0.35\pm0.01$ & 1 & A5 & 36 &&&& $108\pm4$ & $4.7_{-0.06}^{+0.05}$\\
$\rho$\,Cas       & $2.16\pm0.19$ \newline $2.09\pm0.02$ & 1 \newline 37 & \mbox{6000-7500} \newline $7300\pm200$ & 20 \newline 34 & 350--450 \newline \mbox{564--700} \newline 450 \newline - \newline - & $5.42\pm0.3$ \newline - \newline - \newline 5.66 \newline 5.11 & 20 \newline 37 \newline 35 \newline 11 \newline 26 & $653\pm63$ (JDSC) \newline $628\pm24$ (interf.) & $5.86_{-0.19}^{0.13}$ (JDSC)\newline $5.83_{-0.16}^{+0.12}$ (interf.)\\
\hline
\end{tabular}
\tablebib{(1)~\cite{bourges_vizier_2017}; (2)~\cite{rosenzweig_determination_1993}; (3)~\cite{arellano_ferro_comments_1988}; (4)~\cite{kovtyukh_reddenings_2008}; (5)~\cite{luck_parameters_2014}; (6)~\cite{strassmeier_time-series_2014}; (7)~\cite{kovtyukh_accurate_2012}; (8)~\cite{usenko_spectroscopic_2011}; (9)~\cite{arellano_ferro_uvbybeta_1996}; (10)~\cite{forbes_pismis_1994}; (11)~\cite{humphreys_studies_1978}; (12)~\citep{groenewegen_analysing_2020}; (13)~\cite{achmad_photometric_1992}; (14)~\cite{bersier_colour_1996}; (15)~\cite{mantegazza_luminosities_1992}; (16)~\cite{kienzle_pulsating_1998}; (17)~\cite{kourniotis_hd_2025}; (18)~\cite{kipper_optical_2008}; (19)~\cite{sahin_hd_2016}; (20)~\cite{van_genderen_investigation_2025}; (21)~\cite{ikonnikova_multicolor_2018}; (22)~\cite{reddy_spectroscopic_1999}; (23)~\cite{van_genderen_pulsations_2019}; (24)~\cite{hawkins_discovery_1995}; (25)~\cite{oudmaijer_census_2022}; (26)~\cite{klochkova_unity_2019}; (27)~\cite{healy_red_2024}; (28)~\cite{klochkova_optical_2000}; (29)~\cite{hackwell_infrared_1974}; (30)~\cite{anugu_great_2023}; (31)~\citet{van_belle_supergiant_2009}; (32)~\cite{nieuwenhuijzen_hypergiant_2012}; (33)~\cite{kasikov_yellow_2024}; (34)~\cite{israelian_yellow_1999}; (35)~\cite{stothers_yellow_2012}; (36)~\cite{boulon_etude_1963}; (37)~\cite{anugu_chara_2024}.
}
\tablefoot{For V509\,Cas and $\rho$\,Cas we give the radius and luminosity values for both angular diameters: (JDSC) using the angular diameter from \cite{bourges_vizier_2017} and (interf.) using diameters from interferometric studies of \cite{van_belle_supergiant_2009} and \cite{anugu_chara_2024} respectively. \\
\tablefoottext{a}{\cite{groenewegen_analysing_2020} $\log L/L_\odot =5.33$ and $T_\mathrm{eff}=5625$~K, using Stefan-Boltzmann law.}
\tablefoottext{b}{\cite{sahin_hd_2016} $\log L/L_\odot =5.5$ and $T_\mathrm{eff}=7350$~K, using Stefan-Boltzmann law. }
}
\end{table*}

\subsection{Radius}

\begin{figure}[t]
    \centering
    \includegraphics[width=\linewidth]{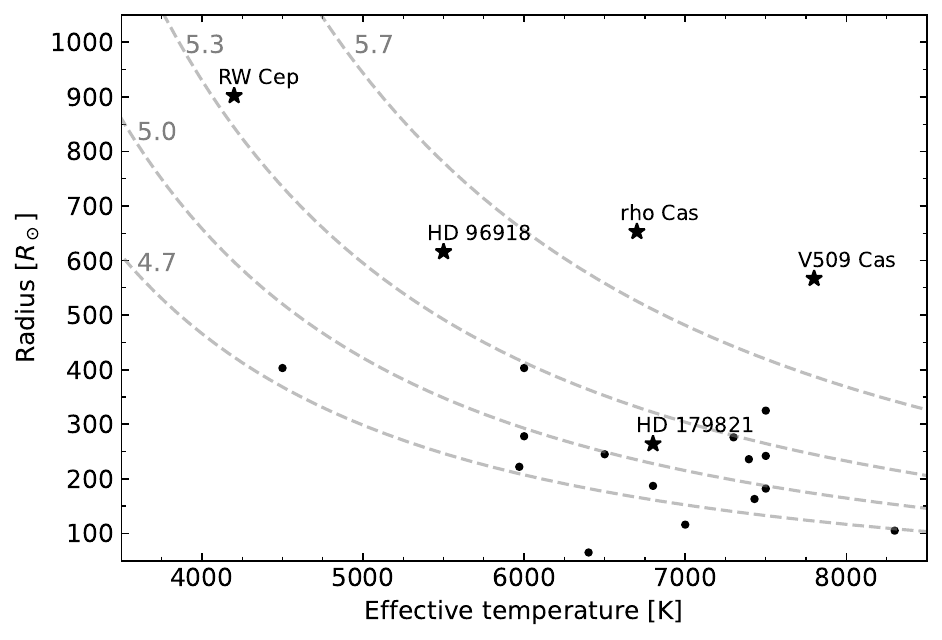}
    \caption{Stellar radii versus \teff. The YHGs are labelled, dashed grey lines indicate constant-luminosity tracks at the indicated $\log L/L_\odot$ values (50\,000, 100\,000, 200\,000, and 500\,000 $L_\odot$.)}
    \label{fig:teff_radius}
\end{figure}

\begin{figure}[ht!]
    \centering
    \includegraphics[width=\linewidth]{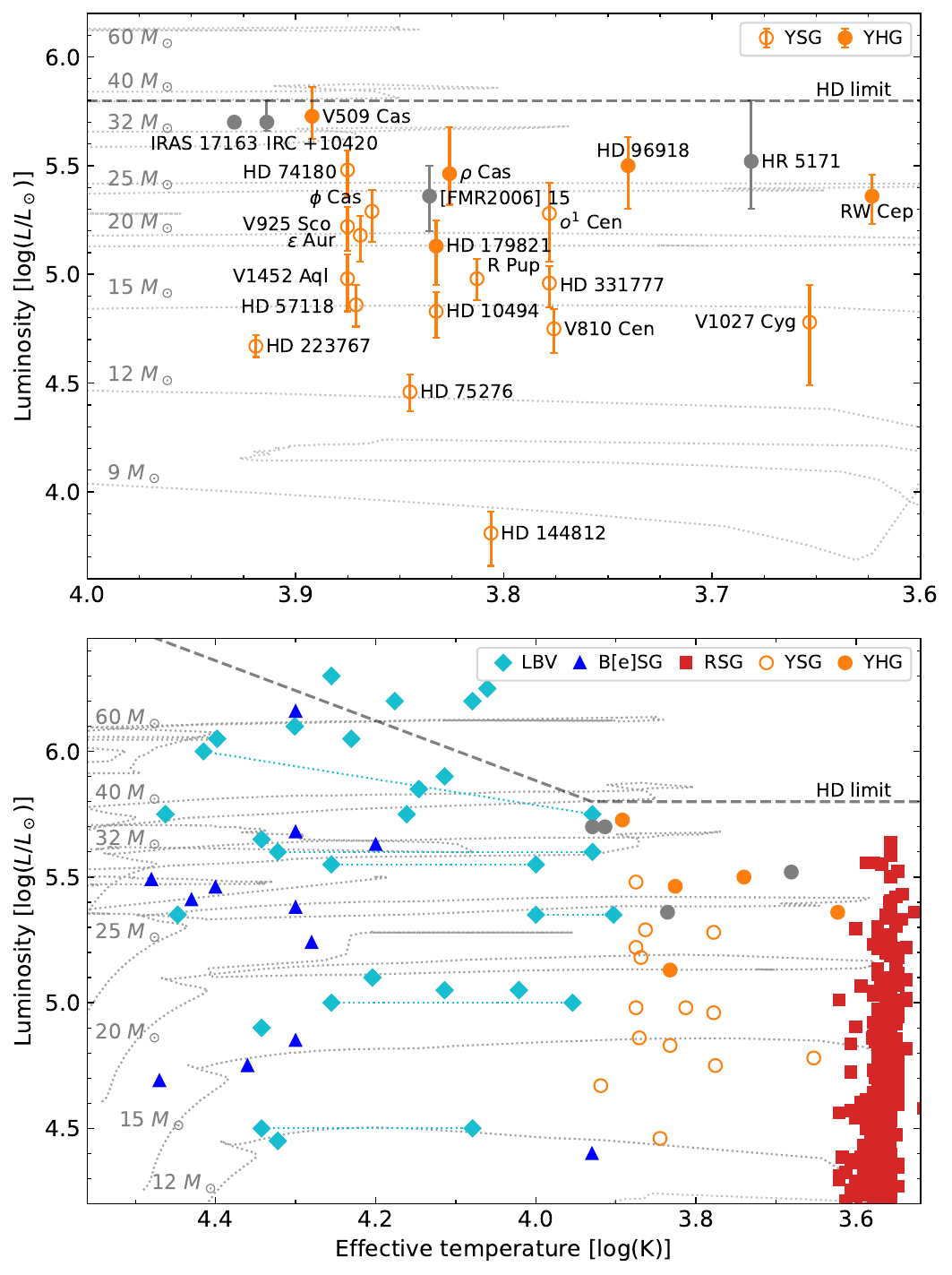}
    \caption{YHGs and YSGs on the HR diagram. YHGs are shown as filled yellow circles, YSGs and candidate YHGs as empty yellow circles. Literature values for four YHGs are shown as grey circles. The HD luminosity limit \citep{humphreys_studies_1978} is marked with a dashed line. Solar-metallicity evolutionary tracks \citep{ekstrom_grids_2012} are shown as light grey dotted lines. Top panel: YHGs and YSGs only. Bottom panel: For comparison, literature data for Galactic RSGs (red squares; \citealt{healy_red_2024}), LBVs (cyan diamonds; \citealt{smith_gaia_2019}), and B[e] Supergiant stars (blue triangles; \citealt{miroshnichenko_be_2025}) are shown.}
    \label{fig:hrd}
\end{figure}

To estimate the stellar radii, we used the limb-darkened angular diameter values from the JMMC Stellar Diameter Catalog (JSDC; \citealt{chelli_pseudomagnitudes_2016,bourges_jmmc_2014,bourges_vizier_2017}). 
Out of the 35 stars in our sample, 20 are included in the JSDC catalogue, providing a homogeneous set of angular diameters based on optical and near-infrared colour indices. Of these, 19 have group-based distance values. HD\,179821 does not have a group-based distance, but has a relatively good \textit{Gaia} parallax and distance from \cite{bailer-jones_estimating_2021}. Using these distances, we calculated the stellar radii ($R/R_\odot = 0.1075~\theta~d$, where $\theta$ is the angular diameter in milliarcseconds and $d$ is the distance in parsecs). The results are listed in Table~\ref{tab:radii_lum}. 

Fig.~\ref{fig:teff_radius} shows the derived radii as a function of effective temperature {\teff} with lines of constant luminosity based on the Stefan-Boltzmann law, $L \propto R^2 T_\mathrm{eff}^4$. The radii of YSGs in our sample range from $100~R_\odot$ to $400~R_\odot$ with uncertainties around 10\%. The radius increases with decreasing effective temperature for stars with similar luminosity. However, the most luminous YHGs -- V509\,Cas and $\rho$\,Cas -- appear to have exceptionally large radii of $\sim600~R_\odot$. To understand this result, we compared the JSDC angular diameter values with angular diameters derived from interferometric measurements. 

For V509\,Cas, the JSDC lists an angular diameter of $1.57\pm0.12$~mas, while the diameter measured by \cite{van_belle_supergiant_2009} is around 20\% smaller, $1.24\pm0.03$~mas. The calculated radii are $567\pm111~R_\odot$ and $449\pm20~R_\odot$, respectively. The latter value is similar to $\sim400~R_\odot$ found by \cite{nieuwenhuijzen_hypergiant_2012}. For $\rho$\,Cas, the JSDC angular diameter is $2.16\pm0.19$~mas, while \cite{anugu_chara_2024} determined an angular limb-darkened diameter of $2.09\pm0.02$~mas from near-infrared interferometry. Using our group-based distance, the resulting radii are $633\pm62~R_\odot$ and $609\pm27R_\odot$, respectively. \cite{van_genderen_investigation_2025} determined the radius of $\rho$\,Cas from its pulsational cycle, and found a quiescent radius of $\sim400~R_\odot$ and an outburst radius of $>700~R_\odot$ at a distance of 2500~pc ($\sim 450~R_\odot$ and $\sim 780 ~R_\odot$ at our group-based distance of 2800~pc). Since the last outburst in 2013, $\rho$\,Cas has been quiescent \citep{van_genderen_investigation_2025}. There is a considerable discrepancy between the radii derived from interferometric and pulsational studies. 

The heterogeneous interferometric studies available for a small number of YSGs and YHGs do not allow us to consistently quantify any systematic offset between the JSDC-inferred angular diameters and those obtained from interferometric measurements. However, comparison with literature values obtained through different methods suggests that the JSDC angular diameters for V509\,Cas and $\rho$\,Cas are likely overestimated. The JSDC empirical relations between angular diameters and photometric indices for luminosity classes I, II and III are primarily calibrated on K- and M-type giants, with very few hotter stars included \citep{chelli_pseudomagnitudes_2016}. For luminous Supergiants in advanced evolutionary states, these relations may not apply. In addition, the colours of YHGs are intrinsically variable \citep{van_genderen_pulsations_2019}, which may contribute to the uncertainty in determining diameters from photometric indices. With limited  and heterogeneous interferometric measurements available, we cannot establish systematic offsets between angular diameters from the JDSC and those obtained from interferometric measurements. Homogeneously derived angular diameters from the spectral energy distribution or interferometric modelling are required to obtain self-consistent radii for the sample stars. 

\subsection{Luminosity}\label{sect:luminosity}

Using the stellar radii derived for 20 stars in the previous section, we estimated the stellar luminosities using the Stefan-Boltzmann law. The resulting luminosities are listed in Table~\ref{tab:radii_lum}, together with literature values for comparison. For studies reporting bolometric magnitudes, we converted them to luminosity using $\log L/L_\odot = 0.4(-M_\mathrm{bol}+4.74)$ \citep{mamajek_iau_2015}. 

The luminosities of YSGs and YHGs are shown in the top panel of Fig.~\ref{fig:hrd}. For V509\,Cas and $\rho$\,Cas, we used the radii from \cite{nieuwenhuijzen_hypergiant_2012} and \cite{van_genderen_investigation_2025} instead of the JSDC values due to the concern of overestimated angular diameters. IRC\,+10420, IRAS 17163-3907, [FMR2006]\,15, and HR\,5171 do not have angular diameters listed in the JDSC, but we adopted their luminosities from the literature for reference. The bottom panel of Fig.~\ref{fig:hrd} shows the high-luminosity region of the HR diagram, including the B[e] Supergiants (B[e]SGs), Luminous Blue Variables (LBVs), and RSGs. The LBV and B[e]SG luminosities have significant error bars, which are not displayed for clarity. 

In the HR diagram, most YHGs occupy the region above $\log L/L_\odot>5.4$, which corresponds to $M_\mathrm{ini}=25-40~M_\odot$. Three stars lie near the Humphreys-Davidson (HD) luminosity limit: IRC\,+10420, IRAS 17163-3907, and V509\,Cas. The YSGs occupy a wider range of luminosities, with the most luminous YSGs partially overlapping with the less luminous YHGs. The luminosities of three YHGs, HD\,179821, RW\,Cep, and HD\,96918, are in good agreement with previous studies. Significant deviations are found for V509\,Cas and $\rho$\,Cas, where using the angular diameters from JDSC results in unexpectedly large radii and unreasonably high luminosities ($\log L/L_\odot >5.8$). Although for $\rho$\,Cas, the large radius is supported by an interferometric study \citep{anugu_chara_2024}. The calculated luminosities for YSGs are generally in very good agreement with literature values. We only find a discrepancy for V810\,Cen, where the previous luminosity estimate of $\log L/L_\odot=5.3$ \citep{kienzle_pulsating_1998} is much higher than our value of $\log L/L_\odot=4.74^{+0.09}_{-0.11}$ due to a revised, lower distance by $\sim$1000~pc. 

\section{Discussion}\label{sect:discussion}

\subsection{Spatial distribution}

Following discussions on the spatial distribution and possible isolation of LBVs and Supergiants with the B[e] phenomenon \citep[e.g.][and references therein]{smith_luminous_2015,aadland_shedding_2018,deman_kinematic_2024,martin_spatial_2025}, \cite{van_genderen_pulsations_2019} proposed that the YHGs $\rho$\,Cas, V509\,Cas, HR\,5171, and HD\,179821 are isolated objects, which could be evidence of an evolutionary connection with LBVs. An explanation for the potential isolation of LBVs is their origin through binary evolution -- either as merger product or as mass gainer in Roche-lobe overflow systems that were kicked out of their original population after their companion exploded as a stripped-envelope supernova \cite{smith_luminous_2015}.

\begin{figure}[]
    \centering
    \includegraphics[width=\linewidth]{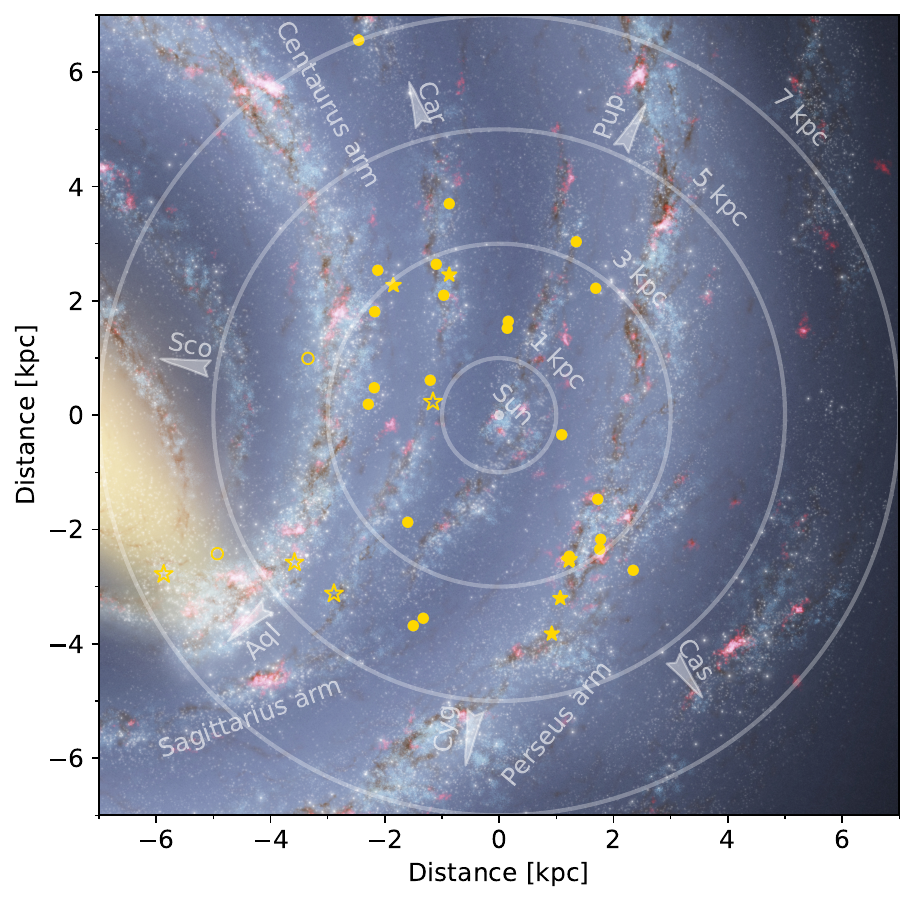}
    \caption{A family portrait. Illustration of the Milky Way{\protect\footnotemark} showing the distribution of YSGs (circles) and YHGs (stars). Objects with group-based distances are marked with filled markers and objects with distances based on \ion{H}{i} or \cite{bailer-jones_estimating_2021} are marked with empty markers. }
    \label{fig:milkyway}
\end{figure}

Fig.~\ref{fig:milkyway} shows the locations of our sample stars projected on a schematic view of the Milky Way. We used the group-based distances where available, otherwise the \ion{H}{I}-based or \cite{bailer-jones_estimating_2021} distances. The stars broadly trace the spiral arms of the Galaxy.
Our results indicate that YHGs are located in heterogeneous environments. Two YHGs are likely cluster members, three are in OB associations, and one is in a star forming region. Four YHGs may be unaffiliated to stellar groups, but two of them are too far to draw a clear conclusion. 

\footnotetext{NASA/JPL-Caltech/R. Hurt (SSC/Caltech)}

In contrast, a majority of the YSGs in our sample are associated with stellar groups. 20 out of 25 YSGs have kinematic properties consistent with membership in open clusters or OB associations. 
For five stars, no clear origin group could be found. In two of these cases, V870\,Sco and IRAS\,14394-6059, the stars are distant and have high extinction. For the other three stars, HD\,12399, $o^1$\,Cen, and HD\,144812, we can identify co-moving stars sharing a similar distance.

Overall, we found that most YSGs are members of young stellar populations. This is consistent with YSGs being younger, pre-RSG objects and still affiliated to their birth environments. The environments of the YHGs are more diverse, and for almost half of them their origin populations remain unclear. This could be the result of different evolutionary histories. If so, there may be multiple ways for a star to become a YHG, and the distance from their birth environments may be a result of binary interactions. 

The number of luminous YSGs found in the Large Magellanic Cloud \citep{martin_census_2023} is more than double the size of our sample and could provide more insight into their surrounding environments. Improved astrometry and distances of Galactic YSGs and YHGs will be provided by the \textit{Gaia} DR4 release\footnote{\url{https://www.cosmos.esa.int/web/gaia/dr4}}, which will include epoch astrometry for all sources. 

\subsection{Runaway stars}\label{sect:runaway}

Around 25--30\% of Galactic O-type stars are runaway stars, and about a quarter of these show bow shocks generated from the interaction between their stellar winds and the interstellar medium \citep{carretero-castrillo_galactic_2023,carretero-castrillo_new_2025}. Studies in the SMC suggest that around 65\% of OB-type field stars are runaways, mainly due to dynamical ejections from clusters \citep[][and references therein]{dorigo_jones_runaway_2020}. Since OB stars are the progenitors of YHGs/YSGs, some stars in our sample may have experienced an ejection event. The YHGs in our sample without clear group affiliation have variable stellar winds and have undergone major mass-loss events \citep[e.g.][]{lobel_high-resolution_2003,oudmaijer_high_1998,koumpia_optical_2020,jura_detached_1999}, setting ideal conditions for bow shock formation in a runaway scenario. However, we found no evidence of a bow shock associated with any YHG or YSG in our sample, either in the literature or in our examination of Wide-Field Infrared Survey Explorer (\textit{WISE}; \citealt{wright_wide-field_2010}) images.

For stars where the \ion{H}{I}-based and group-based distance estimates are in good agreement, a runaway scenario is unlikely, as dynamical perturbations would affect the proper motion, radial velocity, or both, leading to inconsistent distances. An exception is HD\,96918, for which the proper motion aligns with OB associations no.~210/211 \citep{chemel_search_2022}, yielding a group distance of $2623^{+88}_{-83}$~pc, but the \ion{H}{I}-based distance is $\sim$500~pc or $\sim$5000~pc. This discrepancy was already identified by \citet{achmad_photometric_1992}, whose distance estimates of $2400\pm900$~pc and $\sim$2200~pc from independent methods are in good agreement with our group-based distance. If HD\,96918 were a runaway, an apparent alignment with an OB association based on similar proper motion would be unlikely to provide a reliable distance. If HD\,96918 has been dynamically perturbed, its radial velocity could deviate from the \ion{H}{I} velocity, but the effect has not been significant enough to separate it from the stars in the OB association. 

Small offsets in proper motions ($>2\sigma$ of the cluster's mean) are observed for several cluster members: $\phi$\,Cas, HD\,18391, and RW\,Cep have minor proper motion offsets compared to other members of their clusters (Appendix~\ref{app:prop_mot_fig}). HD\,223767 has a projected on-sky offset of $<0.1$~deg of the open cluster King\,12 and shares a similar distance of $\sim2800$~pc, but its proper motion differs from cluster members by more than $3\sigma$. HD\,74180 lies near the open clusters Pismis\,6 and Pismis\,8, both at distances of $\sim1700$~pc and separated by $\sim$0.3~deg on sky, with proper motion differences of $\sim0.5$~\masyr. These cluster may originate from the same molecular cloud \citep{fitzgerald_open_1979}. A proper motion difference of 0.5~\masyr\ corresponds to a velocity difference of $\sim$5~\kms\ at a distance of 2000~pc and $\sim$7~\kms\ at 3000~pc, indicating small kinematical differences originating from modest dynamical interactions. Overall, while we found minor kinematic offsets for several stars, we found no clear evidence of runaway YHGs or YSGs in our sample. 

\subsection{Comments on binarity}

The binary fraction for YSGs is not well constrained. Recently, a multiplicity study in the Small Magellanic Cloud (SMC) suggested that Supergiants of spectral classes B, A, and F were born as effectively single stars or are products of binary mergers \citep{patrick_binarity_2025}. For AF-type Supergiants in the SMC, the binary fraction was estimated to be less than 15\% \citep{patrick_binarity_2025}. However, \citet{ogrady_binary_2024} estimated the binary fraction of YSGs in the Magellanic Clouds at 20--60\%. For reference, O-type stars in the Milky Way have a binary fraction of $\sim$70--90\% \citep{sana_binary_2012,sana_southern_2014}, and the observed binary fraction for RSGs in  the SMC and Large Magellanic Cloud (LMC) is $\sim$15--30\% \citep{neugent_red_2020,dorda_multiplicity_2021,patrick_vlt-flames_2019,patrick_multiplicity_2020,dai_binary_2025}. 

We reprise that a majority of the well-studied YHGs are likely to have binary companions: HR\,5171 is a double or triple system, where one companion could be interacting with the Hypergiant \citep{chesneau_yellow_2014}. 6\,Cas is an A+O-type Supergiant binary \citep{maiz_apellaniz_lucky_2021}. V509\,Cas has a likely B-type companion \citep{lobel_long-term_2013}. IRAS\,17163-3907 has been suggested to have a binary companion \citep{wallstrom_alma_2017}. $\rho$\,Cas, IRC\,+10420, and [FMR2006]\,15 have \textsc{ruwe} values higher than 1.22, which is the threshold commonly used as an indicator for unresolved binarity \citep{castro-ginard_gaia_2024}. Only HD\,179821 appears to be single, as it is unlikely to have a companion more massive than $5~M_\odot$ and a spectral class earlier than B5 \citep{jura_detached_1999}.

Among the YSGs, less information is available, but some confirmed cases of binarity include: $\epsilon$\,Aur is an eclipsing binary \citep{stefanik_epsilon_2010}; HD\,144812 is an interacting binary \citep{kourniotis_hd_2025}; V915\,Sco has a Wolf-Rayet companion \citep{andrews_hr_1977}; V810\,Cen is a spectroscopic binary \citep{kienzle_pulsating_1998}; $\phi$\,Cas and HD\,57118 have been identified as binaries \citep{burki_nineteen_1983}; and HD\,96918, HD\,74180, and IRAS\,18357-0604 have high \textsc{ruwe} values. 

Overall, about 50\% of the YHGs and YSGs in our sample show direct or indirect signs of binarity, with a firm lower limit of 22\% (eight confirmed binary systems). We selected our sample based on luminosity estimates in the literature and did not consider any binarity indicators. Nevertheless, these factors are correlated through the contribution of a hot companion to the overall luminosity of the system, which may introduce a bias towards identifying binaries.

\section{Conclusions}

In this study, we provided a homogeneous and consistent determination of distances for Galactic YSGs and YHGs. 
Due to the uncertainties in the \textit{Gaia} parallaxes, we explored an indirect method of distance determination through membership identification with nearby stellar groups -- clusters or OB associations. We compared the proper motions of YHGs and YSGs with those of clusters \citep{hunt_improving_2024} and OB associations \citep{chemel_search_2022,melnik_kinematics_2017} in the surrounding sky region. Using the more reliable parallaxes of co-moving stars, we calculated ``group-based'' distances. We validated this method against the large-scale Galactic kinematics by comparing the stellar systemic radial velocities with the Galactic \ion{H}{I} velocity map \citep{soding_spatially_2025}. Our ``\ion{H}{I}-based''distances agree well with previous studies, where distances were derived from comparison with the Galactic rotation curve. The two independent methods show good agreement for majority of the targets.

We determined membership in a cluster or OB association for 20 YSGs. Five YSGs remain without a clearly identified stellar group, two of them are likely too distant and reddened for the method. 
The stellar environments of YHGs are more varied:
\begin{itemize}
    \item Stellar cluster: RW\,Cep, [FMR2006]\,15
    \item OB association: 6\,Cas, V509\,Cas, HD\,96918
    \item Star forming region: HR\,5171
    \item \raggedright Unaffiliated: $\rho$\,Cas, IRAS\,17163-3907, IRC\,+10420, HD\,179821
\end{itemize}

We derived luminosities for 15 YSGs and 5 YHGs by combining our distance estimates with {\teff} values from the literature and angular diameters from the JSDC catalogue. Our results are generally in good agreement with previous luminosity estimates. The upcoming \textit{Gaia} DR4 will improve the astrometric measurements and allow for further studies of stellar environments surrounding YHGs and YSGs. A homogeneous interferometric survey of Galactic YSGs and YHGs would be valuable to better constrain the radii and luminosities, determine the binary fraction, and study the circumstellar environments. Future work will address the circumstellar environments and spectral energy distributions of YSGs and YHGs. 

\begin{acknowledgements}
We would like to thank Willem-Jan de Wit, Evgenia Koumpia, and Rain Kipper for insightful discussions. 
This work has made use of data from the European Space Agency (ESA) mission
{\it Gaia} (\url{https://www.cosmos.esa.int/gaia}), processed by the {\it Gaia}
Data Processing and Analysis Consortium (DPAC,
\url{https://www.cosmos.esa.int/web/gaia/dpac/consortium}). Funding for the DPAC
has been provided by national institutions, in particular the institutions
participating in the {\it Gaia} Multilateral Agreement.

This research has made use of the Jean-Marie Mariotti Center JSDC catalogue, which involves the JMDC catalogue.
JSDC available at \url{http://www.jmmc.fr/jsdc}.

Part of this work was supported by the Estonian Research Council grant PRG 2159.

This research has made use of the VizieR catalogue access tool, CDS, Strasbourg, France.

\end{acknowledgements}

\bibliographystyle{aa}
\bibliography{references.bib}

\clearpage
\begin{appendix}
\section{Notes on individual targets}\label{sect:distance_comments_stars}

In this section, we provide a brief overview of the analysis performed for the individual targets. Unless stated otherwise, the parameters used to select nearby stars are the same as those described in Sect.~\ref{sect:distance}. 

\textit{$\phi$ Cas}. The \textit{Gaia} DR3 parallax is low-quality ($\varpi/\sigma_\varpi=2.6$). Its on-sky location coincides with the cluster NGC\,457 ($2800\pm9$~pc; \citealt{hunt_improving_2024}) and it has been included as a cluster member \citep{arellano_ferro_colour_1990,rastorguev_statistical_1999}. Its proper motion does not support cluster membership, but the errors are relatively large ($\sim$0.06~\masyr). We applied a $0.4$~{\masyr} proper motion cut around $\phi$\,Cas' value on stars within 10{\arcmin} of the target. We calculated a group distance of $2920^{+114}_{-113}$~pc, consistent with membership in cluster NGC\,457. The radial velocity of $\phi$\,Cas from \textit{Gaia} DR3 is $-19.2\pm3$~\kms, while other studies report values near $-28$~{\kms} \citep{arellano_ferro_comments_1988,rosenzweig_determination_1993}. The last is similar to the cluster radial velocity of $-34$~{\kms} \citep{rastorguev_statistical_1999}. Using the radial velocity $-28$~\kms, we found a \ion{H}{I}-based distance in the range of 1000--1800~pc, much closer than the cluster. Both the star's and the cluster's radial velocities are inconsistent with the \ion{H}{I} radial velocity at 2800~pc. We found the cluster NGC\,457 to be the most likely origin population.

\textit{\object{HD 10494}}. The \textit{Gaia} DR3 parallax is high-quality ($\varpi/\sigma_\varpi=19.8$), resulting in a distance of $2823^{+142}_{-114}$~pc \citep{bailer-jones_estimating_2021}. The on-sky location of HD\,10494 coincides with the cluster NGC\,654 ($2815\pm11$~pc; \citealt{hunt_improving_2024}) and the star has been included as a cluster member \citep{arellano_ferro_colour_1990,rastorguev_statistical_1999} and a member of the OB association Cas\,OB8 in the same sky region \citep{melnik_kinematics_2017}. Its proper motion is in good agreement ($2\sigma$) with cluster membership. The proper motion also aligns very well with the OB association no.~150 at a distance of 2626~pc \citep{chemel_search_2022}. We applied a $0.2$~{\masyr} proper motion cut around HD\,10494's value on stars within 10{\arcmin} of the target. We calculated a group distance of $2784^{+57}_{-61}$~pc, very similar to the \cite{bailer-jones_estimating_2021} distance. The radial velocity from \textit{Gaia} DR3, $-30.6$~{\kms}, agrees well with the cluster velocity of $-33.8$~{\kms} \citep{rastorguev_statistical_1999}. We found a \ion{H}{I}-based distance in the range of 1000--1800~pc, with a similar discrepancy as for $\phi$\,Cas. The two stars lie within 5$\degree$ of each other on the sky, which corresponds to a physical distance of $\sim$270~pc at their group-based distances. 

\textit{\object{HD 12399}}. The \textit{Gaia} DR3 parallax is high-quality ($\varpi/\sigma_\varpi=14.3$), resulting in a distance of $3792^{+240}_{-192}$~pc \citep{bailer-jones_estimating_2021}. Its on-sky location and proper motion are not aligned with any known cluster or OB association. We applied a $0.3$~{\masyr} proper motion cut around HD\,12399's value on stars within 10{\arcmin} of the target. Most stars within the cut have distances >3500~pc, but their parallaxes are low quality. There are six foreground stars at 2200--2700~pc, but given the high-quality parallax of HD\,12399, the farther population is more likely. Among the more distant stars, only four stars have $\varpi/\sigma_\varpi>5$. We calculated a group distance of $3564^{+320}_{-266}$~pc. Using the \textit{Gaia} DR3 radial velocity  of $-44\pm8$~\kms, we found a \ion{H}{I}-based distance in the range of 2000--3200~pc. 

\textit{\object{HD 18391}}. The \textit{Gaia} DR3 parallax is high-quality ($\varpi/\sigma_\varpi=18.6$), resulting in a distance of $2263^{+113}_{-93}$ \citep{bailer-jones_estimating_2021}. The on-sky location of HD\,18391 coincides with the sparse open cluster SAI\,25 ($2252\pm11$~pc; \citealt{hunt_improving_2024}), \citet{turner_cepheid_2009} described its home cluster as `Anonymous' and determined a shorter distance of $1661\pm73$~pc. Its proper motion is in agreement ($3\sigma$) with membership in SAI\,25. The proper motion also aligns very well with the OB association no.~143 at a distance of 2262~pc \citep{chemel_search_2022}. We applied a $0.3$~{\masyr} proper motion cut around HD\,18391's value on stars within 10{\arcmin} of the target. We calculated a group distance of $2242^{+49}_{-52}$~pc. Using the \textit{Gaia} DR3 radial velocity of $-39\pm3$~\kms, we found a \ion{H}{I}-based distance in the range of 2000--2800~pc.

\textit{\object{$\epsilon$ Aur}}. The \textit{Gaia} DR3 parallax is medium-quality ($\varpi/\sigma_\varpi=5.5$) resulting in a distance of $1056^{+218}_{-182}$~pc \citep{bailer-jones_estimating_2021}. It has a high \textsc{ruwe} value of 1.71 and large proper motion uncertainties ($\sim0.2$~\masyr). Its on-sky location and proper motion are not aligned with any known cluster. It has been identified as a member of the Aur\,OB1 association \citep{arellano_ferro_colour_1990,melnik_kinematics_2017}. On-sky, it is close to OB association no.~147 at a distance of 1154~pc \citep{chemel_search_2022}, though its proper motion differs by $\sim0.6$~{\masyr} from the association members. But given the large errors, we consider the proper motions similar enough. Within 20{\arcmin} of the target we found most stars at distances >2000~pc. There are very few hot stars at $\sim$1000~pc and they are scattered in proper motions. We applied a very wide $1.6$~{\masyr} proper motion cut around $\epsilon$\,Aur's value and found only four stars. The group-based distance is $1127\pm31$~pc. Using the radial velocity of $-2.3$~{\kms} \citep{stefanik_epsilon_2010}, we found a \ion{H}{I}-based distance of $\sim$700~pc, consistent with the closer group distance. 

\textit{\object{HD 57118}}. The \textit{Gaia} DR3 parallax is high-quality ($\varpi/\sigma_\varpi=18$), resulting in a distance of $2759^{+116}_{-138}$~pc. It is a member of the OB association no.~25 at a distance of 2524~pc \citep{chemel_search_2022}. We applied a $1$~{\masyr} proper motion cut around HD\,57118's value on stars within 10{\arcmin} of the target. We calculated a group distance of $2790^{+88}_{-93}$~pc. Using the \textit{Gaia} DR3 radial velocity $61.2\pm1.4$~\kms, we found a \ion{H}{I}-based distance of $\sim$3300~pc. 

\textit{R Pup}. The \textit{Gaia} DR3 parallax is high-quality ($\varpi/\sigma_\varpi=13.5$), resulting in a distance of $3958^{+291}_{-228}$~pc. The on-sky location of R\,Pup coincides with the cluster NGC\,2439 ($3350\pm13$~pc; \citealt{hunt_improving_2024}) of which it is a member \citep{arellano_ferro_colour_1990,white_ubv_1975}. Its proper motion is in good agreement ($1\sigma$) with cluster membership. The proper motion also aligns very well with the OB association no.~68 at a distance of 3237~pc \citep{chemel_search_2022}. We applied a $0.1$~{\masyr} proper motion cut around R\,Pup's value on stars within 10{\arcmin} of the target. We calculated a group distance of $3331^{+77}_{-80}$~pc. Using the radial velocity of  $62.5$~{\kms} \citep{balona_radial_1982}, we found a \ion{H}{I}-based distance of $\sim$4000~pc. 

\textit{HD 74180}. The \textit{Gaia} DR3 parallax is medium-quality ($\varpi/\sigma_\varpi=4.3$), resulting in a distance of $2532^{+600}_{-397}$~pc. It has high \textsc{ruwe} value of 2.56 and large proper motion uncertainties ($\sim$0.1~\masyr). It has been identified as a member of the Vel\,OB1 association \citep{melnik_kinematics_2017,humphreys_studies_1978,reed_vela_2000}.
The proper motion aligns with the OB association no.~107 at a distance of 1935~pc \citep{chemel_search_2022}. 
The star has been proposed to be affiliated with the region of the open cluster NGC\,2645 (Pismis\,6; \citealt{forbes_pismis_1994}) at a distance of $1720\pm5$~pc \citep{hunt_improving_2024}, although it is not considered a member \citep{aidelman_open_2015}. Its proper motion has a $\sim$0.5~mas~yr$^{-1}$ offset from the members of NGC\,2645. The cluster radial velocity 
of $20.8\pm3$~{\kms} \citep{rastorguev_statistical_1999} is similar to the radial velocity of HD\,74180 of $25\pm3$~{\kms} \citep{forbes_pismis_1994}. We applied a $0.5$~{\masyr} proper motion cut around HD\,74180's value on stars within 20{\arcmin} of the target. Hot stars with similar proper motion to HD\,74180 are located at cluster distance. Because its proper motion does not align clearly with the cluster, we do not count it as a member, but we found it to be affiliated with the same region, in agreement with \cite{forbes_pismis_1994}. We calculated a group distance of $1657_{-34}^{+33}$~pc. Using the radial velocity of $25$~{\kms} \citep{forbes_pismis_1994}, we found a \ion{H}{I}-based distance in the range of 1500--2000~pc. 

\textit{\object{HD 75276}}. The \textit{Gaia} DR3 parallax is high-quality ($\varpi/\sigma_\varpi=19$), resulting in a distance of $1443^{+74}_{-57}$~pc. It has been identified as a member of the Vel\,OB1 association \citep{melnik_kinematics_2017,reed_vela_2000}. The proper motion aligns with the OB association no.~106 at a distance of 1447~pc or no.~107 at a distance of 1935~pc \citep{chemel_search_2022}. In the 10{\arcmin} field of view, there is a low number of hot stars with similar proper motions and parallax, and there is a significant distant population at $\sim$3000~pc. We applied a $0.7$~{\masyr} proper motion cut around HD\,75276's value on stars within 10{\arcmin} of the target and excluded the background stars. We calculated a group distance of $1538\pm24$~pc. Using the \textit{Gaia} DR3 radial velocity of $\sim$26~\kms, we found a \ion{H}{I}-based distance of $\sim$1500~pc. \cite{de_medeiros_catalog_2002} suggested a slightly lower velocity of 21~{\kms} and \cite{reed_vela_2000} derived a higher velocity of $\sim$32~\kms. These velocities decrease or increase the distance estimate by a couple of hundred parsecs.

\textit{\object{V709 Car}}. The \textit{Gaia} DR3 parallax is high-quality ($\varpi/\sigma_\varpi=8.1$), resulting in a distance of $4006^{+397}_{-376}$. We did not find alignment with any known cluster or OB association. In the 10{\arcmin} field of view there is a significant population of hot stars at $\sim$4000~pc with proper motions similar to V709\,Car, but their parallax errors are large. We applied a $0.25$~{\masyr} proper motion cut around V709\,Car's value, excluded likely foreground stars at $\sim$2000~pc, and relaxed the parallax quality criterion to $\varpi/\sigma_\varpi=3$. We calculated a group distance of $3817^{+315}_{-271}$~pc. V709\,Car is a binary system with a period of 323 days, its radial velocity varies between $\sim6-18$~{\kms} \citep{maas_study_2003}. Using an average radial velocity of 12~\kms, we found a \ion{H}{I}-based distance of $\sim$3600~pc.

\textit{\object{TYC 8958-479-1}}. The \textit{Gaia} DR3 parallax is medium-quality ($\varpi/\sigma_\varpi=3$), resulting in a distance of $8783^{+2122}_{-1330}$~pc. \cite{maiz_apellaniz_barba_2025} identified the star as a member of the open cluster Barbá\,2 at a distance of $7390^{+650}_{-550}$. Its proper motion also aligns with the OB association no.~8 \citep{chemel_search_2022} at a distance of 6917~pc. The Barbá\,2 cluster is not included in the \citet{hunt_improving_2024} catalogue, thus we do not include cluster membership in the identified cluster column of Table~\ref{tab:distances}. We applied a $0.1$~{\masyr} proper motion cut around TYC\,8958-479-1's value on stars within 10{\arcmin} of the target and identified four hot stars at this large distance with reliable parallaxes. We calculated a group distance of $7025^{+687}_{-572}$~pc. We are unaware of any radial velocity measurements to compare with the \ion{H}{I} kinematics. 

\textit{\object{HD 96918}}. The \textit{Gaia} DR3 parallax is low-quality ($\varpi/\sigma_\varpi=2.2$). The star has a high \textsc{ruwe} value of 1.99 and large proper motion uncertainties ($\sim0.1$~\masyr). It has been identified as a member of the Car\,OB2 association \citep{melnik_kinematics_2017,humphreys_studies_1978}. On the sky, it is located at the edge of the open cluster NGC\,3532 (distance 472~pc; \citealt{hunt_improving_2024}), but its proper motion differs from the cluster members by $\sim$4~\masyr, and thus is incompatible with membership. We found a relatively good alignment in proper motion with the OB associations 210 and 211 at distances of 2362~pc and 2450~pc \citep{chemel_search_2022}, but they are $\sim$1{\degree} away on the sky. We applied a $0.4$~{\masyr} proper motion cut around HD\,96918's value on stars within 10{\arcmin} of the target. We found stellar populations at $\sim$2500~pc and at $\sim$4000~pc. Based on previous distance estimates in the literature -- 1900~pc \citep{luck_parameters_2014}, 2000~pc \citep{groenewegen_analysing_2020}, and 2700~pc \citep{achmad_photometric_1992} -- HD\,96918 is more likely part of the closer population. We calculated a group distance of $2623_{-83}^{+88}$~pc. Using the radial velocity of 5.6~{\kms} \citep{balona_radial_1982}, we found a \ion{H}{I}-based distance of $\sim$5000~pc or $\sim500$~pc, similar to the result found by \cite{achmad_photometric_1992}. This discrepancy is addressed in Sect.~\ref{sect:discussion_distance}. 

\textit{\object{$o^1$ Cen}}. The \textit{Gaia} DR3 parallax is medium-quality ($\varpi/\sigma_\varpi=4.43$), resulting in a distance of $3783^{+3197}_{-928}$~pc.  We did not find alignment with any known cluster or OB association. We applied a $0.6$~{\masyr} proper motion cut around $o^1$\,Cen's value on stars within 20{\arcmin} of the target. We found a population at $\sim$3000~pc with similar proper motion to $o^1$\,Cen, no foreground group, but a significant background population at >5000~pc. We calculated a group distance of $2880\pm100$~pc. Using the radial velocity of $-21$~{\kms} \citep{gontcharov_pulkovo_2006}, we found a \ion{H}{I}-based distance of $\sim$2800~pc. 

\textit{\object{V810 Cen}}. The \textit{Gaia} DR3 parallax is low-quality ($\varpi/\sigma_\varpi=1.7$). The star has large proper motion uncertainties ($\sim$0.1~mas~yr$^{-1}$) likely due to binarity. It has been identified as a member of the open cluster Stock\,14 \citep{kovtyukh_accurate_2010} at a distance of $2340\pm9$~pc \citep{hunt_improving_2024} and also as a background star \citep{kienzle_pulsating_1998}. On-sky, it is located within the cluster Stock\,14. The \cite{hunt_improving_2024} catalogue lists two other open clusters in the same sky region: Theia\,2727 ($3655_{-50}^{+52}$~pc) and Lynga\,15 ($1646\pm8$~pc). The proper motion does not align with any of the clusters. Within a 10{\arcmin} field of view, we applied a proper motion cut with a radius of $0.4$~{\masyr} around V810\,Cen and found stellar populations at the distance of Stock\,14 and a farther beyond 3000~pc.\cite{kienzle_pulsating_1998} assumed, based on previous spectroscopic analysis, that the blue companion has a B0-B1 Iab-Ib spectral type and thus an absolute magnitude $M_V\simeq-6$~mag, suggesting a distance greater than 3000~pc. However, in a more recent analysis \citep{wegner_absolute_2006}, these supergiant spectral types correspond to $-5.22\leq M_V \leq -4.65$~mag, consistent with $M_V=-4.6$~mag found by \cite{parsons_hr_1981} who assumed a cluster distance of 2600~pc. Another indication of the closer distance is given by interstellar extinction. E($B-V$) is in the range of 0.24--0.3 for both Stock\,14 and V810\,Cen \citep{parsons_hr_1981,kienzle_pulsating_1998}. Within 10{\arcmin} of the target we found low extinction for stars at a distance of 2200--2500~pc (monochromatic extinction from \textit{Gaia} $0.5 \leq A_0 \leq 1$), and much higher extinction for stars at distances of 3000--4000~pc, with $A_0$ increasing from 2 to 4 or even 5. In Wide-Field Infrared Survey Explorer (\textit{WISE}; \citealt{wright_wide-field_2010}) band 3 (12~$\mu$m) and band 4 (22~$\mu$m) images, a dense structure can be seen at the location of Stock\,14, which corresponds to the \ion{H}{II} region \object{GAL 295.14-00.63} \citep{anderson_wise_2014}. We extracted the \ion{H}{II} number density from the map of \cite{soding_spatially_2025} in the line-of-sight towards V810\,Cen, and found that it increases beyond 2000~pc and peaks at $\sim$2800~pc. This suggests that the \ion{H}{II} region lies beyond the cluster Stock\,14, causing enhanced extinction for its background stars. Due to the low reddening of V810\,Cen, it cannot be located within this region and must be part of a closer population. We calculated a group distance of $2330_{-65}^{+61}$~pc for the closer population. We also found a nearby OB association, no.~205 at a distance of 2347~pc \cite{chemel_search_2022}, that overlaps with the Cru\,OB1 association \citep{melnik_kinematics_2017}. The proper motion of the stars in the OB association are a better match for V810\,Cen than the Stock\,14 stars, but most of its members in the \cite{chemel_search_2022} catalogue are more than 1~deg away on sky. Using the radial velocity of $v_\mathrm{rad}=-17$~{\kms} \citep{kienzle_pulsating_1998}, we found a \ion{H}{I}-based distance in the range of 2400--3200~pc.

\textit{\object{HR 5171}}. The \textit{Gaia} DR3 parallax is medium-quality ($\varpi/\sigma_\varpi=4.79$), resulting in a distance of $3601^{+649}_{-539}$~pc. The YHG HR\,5171 has been associated with the star-forming region Gum\,48d/R\,80 \citep[][and references therein]{karr_gum_2009}. Its on-sky location and proper motion are not aligned with any known cluster or OB association. We applied a $0.4$~{\masyr} proper motion cut around HR\,5171's value on stars within 10{\arcmin} of the target. We found a relatively smooth distribution of stellar populations between 2000 and 3500~pc with two population peaks at distances 2300~pc and 2900~pc. Stars at farther distances have high parallax uncertainties, likely due to increased extinction in the \ion{H}{II} region Gum\,48d (R\,80) located at a distance of 2900~pc \citep{melnik_internal_2020} or 3600~pc \citep{karr_gum_2009}. Most stars with proper motions similar to HR\,5171 belong to the 2900~pc population. We calculated a group distance of $2953_{-96}^{+92}$~pc, which we adopted for HR\,5171, but the distance could be higher, as the hot stars with similar proper motions at farther distances are not taken into account because of high parallax uncertainties. The group-based distance of the foreground population is $2297^{+63}_{-65}$~pc, which could be considered a lower limit for the distance for HR\,5171. HR\,5171A has a companion HR\,5171B, which is a B0Ib spectral type supergiant \citep{humphreys_high-luminosity_1971,karr_gum_2009} with a high-quality parallax value ($\varpi/\sigma_\varpi=19$) and a distance of $2879^{+148}_{-134}$~pc \citep{bailer-jones_estimating_2021}. Using the radial velocity of $v_\mathrm{rad}=40$~{\kms} \citep{balona_radial_1982,humphreys_high-luminosity_1971}, we found a \ion{H}{I}-based distance of $\sim$3500~pc (see Fig.~\ref{fig:HIyhgs}). Thus, HR\,5171 is at a distance of 2900--3600~pc for four reasons: 1) good agreement with \ion{H}{I} kinematics and Galactic rotation curve; 2) stellar group at that distance; 3) agreement with the \textit{Gaia} parallax-based distance of $3601^{+649}_{-539}$ ($\varpi/\sigma_\varpi\approx5$; \citealt{bailer-jones_estimating_2021}; 4) the companion has a similar distance with high-quality quality parallax.

\textit{UCAC2 4867478}. The \textit{Gaia} DR3 parallax is medium-quality ($\varpi/\sigma_\varpi=6.2$), resulting in a distance of $3327^{+465}_{-331}$~pc. We did not find alignment with any known cluster or OB association. We applied a $0.2$~{\masyr} proper motion cut around UCAC2\,4867478's value on stars within 10{\arcmin} of the target. We found a group of hot stars with very similar proper motions and parallaxes. We relaxed the parallax quality criterion to $\varpi/\sigma_\varpi=3$ and calculated a group distance of $3322^{+351}_{-314}$~pc. Using the \textit{Gaia} DR3 radial velocity of $-41\pm4$~{\kms}, we found a \ion{H}{I}-based distance of $\sim$3600~pc.

\textit{\object{IRAS 14394-6059}}. The \textit{Gaia} DR3 parallax is medium-quality ($\varpi/\sigma_\varpi=5.3$, resulting in a distance of $5450^{+770}_{-686}$~pc \citep{bailer-jones_estimating_2021}. We did not find alignment with any known cluster or OB association. Within a 10{\arcmin} field of view, we tried different proper motion cuts and found hot stars scattered over a wide range of distances from 2000~pc to 5000~pc. The parallaxes of distant stars have very high uncertainties and we have no way to disentangle the populations. We are not aware of any radial velocity measurements to derive \ion{H}{I}-based distance. If the \textit{Gaia} parallax is reliable, it is potentially a luminous YSG.

\textit{\object{CD-59 5594}}. The \textit{Gaia} DR3 parallax is high-quality ($\varpi/\sigma_\varpi=10.33$), resulting in a distance of $3739^{+325}_{-279}$~pc \citep{bailer-jones_estimating_2021}. The star's proper motion aligns with the cluster Pismis\,21 at a distance of $2906\pm24$~pc \citep{hunt_improving_2024} and with the OB association no.~47 at 2958~pc \citep{chemel_search_2022}. We applied a $0.15$~{\masyr} proper motion cut around CD-59\,5594's value on stars within 10{\arcmin} of the target. We calculated a group distance of $2849\pm77$~pc. Using the \textit{Gaia} DR3 radial velocity of $-36\pm4$~{\kms}, we found a \ion{H}{I}-based distance of $\sim$2500~pc.

\textit{HD 144812}. The \textit{Gaia} DR3 parallax is high-quality ($\varpi/\sigma_\varpi=21.55$), resulting in a distance of $1352^{+66}_{-58}$~pc \citep{bailer-jones_estimating_2021}. However, the \textsc{ruwe} value is high (1.37). \cite{kourniotis_hd_2025} proposed that HD\,144812 has a companion of spectral class B2.5 or hotter. The proper motion of HD\,144812 does not align with any known cluster or OB association. Within a 10{\arcmin} field of view, we found very few hot stars. We applied a $0.5$~{\masyr} proper motion cut around HD\,144812's value and found only four stars with $T_\mathrm{eff}>7500$~K: two with distances $\sim$1300~pc and two with distances $\sim$1500~pc. The group-based distance estimate for these four stars is $1454^{+52}_{-51}$~pc. Increasing the cut radius did not show any farther population with similar kinematics. We are unaware of any radial velocity measurements to calculate a \ion{H}{I}-based distance. We explored the effect of the high \textsc{ruwe} value on the distance uncertainty of HD\,144812. \cite{el-badry_how_2025} derived a function that describes the parallax uncertainty inflation factor, $f$, which takes the higher \textsc{ruwe} values into account. To calculate $f$, we used the analytic function given in Eq.~3 of \cite{el-badry_how_2025}. In addition to the \textsc{ruwe} value, the inflation factor depends on the parallax itself and an empirical along-scan measurement, $\sigma_\eta$, which is a function of $G$-magnitude, \cite[see Sect.~2 in][for a detailed description]{el-badry_generative_2024}. HD\,144812 has $G=7.9$~mag, and based on Fig.~3 in \cite{holl_gaia_2023}, we adopted $\sigma_\eta \approx 0.2$~mas. We calculated an inflation factor of $f=2.3$, resulting in a corrected parallax uncertainty of $\sigma_\varpi=0.0773$. For the parallax itself, we applied the zero-point correction \citep{lindegren_gaia_2021}. The resulting corrected parallax is $\varpi = 0.7497 \pm 0.0773$~mas. Even after inflating the parallax error, $\varpi/\sigma_\varpi=9.34$, indicating a parallax uncertainty of $\sim 10 \%$ and confirming that the \textit{Gaia} parallax of HD\,144812 is reasonably reliable. Based on the corrected parallax we derived a geometric distance of $1374_{-132}^{+169}$~pc, which we adopt for HD\,144812. 

\textit{\object{V870 Sco}}. The \textit{Gaia} DR3 parallax is medium-quality ($\varpi/\sigma_\varpi=3.87$), resulting in a distance of $3945^{+940}_{-642}$~pc \citep{bailer-jones_estimating_2021}. On the sky, it is located in the open cluster NGC\,6231 (distance $1553\pm2$~pc; \citealt{hunt_improving_2024}), but V870\,Sco is significantly more reddened than the cluster members ($A_V \approx 10.0$) and has been proposed as a background star \citep{damiani_chandra_2016}. This might be a result of an optically thick circumstellar shell or that the star lies behind the cluster, as its radial velocity is different from the cluster members \citep{robinson_nature_1973,damiani_chandra_2016}. Its \textit{Gaia} proper motion differs by 1.5~\masyr from the cluster members. Detecting a background population is challenging due to the high extinction behind the cluster. We tested 10\arcmin\ and 20\arcmin\ fields of view, but the results remain inconclusive. Thus, we are not able to align V870\,Sco with any known stellar cluster or association. Using the \textit{Gaia} DR3 radial velocity of $-43\pm5$~{\kms}, we found a \ion{H}{I}-based distance of $\sim$3500~pc. The radial velocity of $-53$~{\kms} from \cite{herbig_highly_1972} results in a distance of $\sim$4000~pc, both in agreement with the \cite{bailer-jones_estimating_2021} distance.

\textit{\object{V915 Sco}}. The \textit{Gaia} DR3 parallax is high-quality ($\varpi/\sigma_\varpi=10.44$), resulting in a distance of $1809_{-182}^{+191}$~pc \citep{bailer-jones_estimating_2021}. V915\,Sco is a luminous YSG (HD 155603A) with a Wolf-Rayet WN companion WR~85 (HD 155603B; \citealt{andrews_hr_1977}). On the sky, it is located in the open cluster HSC\,2870 at a distance of $2350\pm25$~pc \citep{hunt_improving_2024} and its proper motion aligns with it (1$\sigma$). The proper motion also aligns with the OB association no.~82 at a distance of 2250~pc \citep{chemel_search_2022}. We applied a $0.2$~{\masyr} proper motion cut around V915\,Sco's value on stars within 10{\arcmin} of the target. We calculated a group distance of $2253_{-56}^{+57}$~pc. Using the radial velocity of $-6$~\kms \citep{andrews_hr_1977}, we found a \ion{H}{I}-based distance in the range of 1200--2000~pc. 

\textit{\object{IRAS 17163-3907}}. The \textit{Gaia} DR3 parallax is low-quality ($\varpi/\sigma_\varpi=1.95$). According to literature, IRAS 17163-3907 could be a member of the star forming region RCW~122 at a distance of $3380^{+330}_{-270}$~pc \citep{wu_trigonometric_2012} or 5000~pc \citep{arnal_molecular_2008}. The radial velocity of the molecular cloud has been measured at $v_\mathrm{LSR}=-15$~\kms, where gas with velocity from $-23$ to $-8$~\kms is expected to be physically associated \citep{arnal_molecular_2008}. These values differ from the star's systemic velocity $v_\mathrm{LSR}=18$~\kms \citep{wallstrom_alma_2017}. Calculations using luminosity and the star's high visual extinction suggest a 1000~pc distance \citep{wallstrom_investigating_2015} and the distance derived from \textit{Gaia} DR2 parallax, $1200^{+400}_{-200}$~pc, has been used by \cite{koumpia_optical_2020}. A distance between 3600~pc and 4700~pc was suggested by \cite{lagadec_double_2011}. We found no alignment with any known stellar cluster or OB association, but note that due to the high extinction in this region, any stars farther than 3000~pc could be too faint for reliable parallaxes. Using the $v_\mathrm{LSR}=18$~\kms \citep{wallstrom_alma_2017} and an LSR correction of $\sim$8~\kms, we found a \ion{H}{I}-based distance of $\sim$1000~pc (see Fig.~\ref{fig:HIyhgs}). 

\textit{\object{V925 Sco}}. The \textit{Gaia} DR3 parallax is high-quality ($\varpi/\sigma_\varpi=12.76$), resulting in a distance of $3006^{+192}_{-166}$~pc. It has been identified as a member of Trumpler~27 \citep{moffat_trumpler_1977}, although this cluster might be a collection of stellar populations extending from 1500 to 5000~pc rather than a compact group \citep{perren_photometric_2012}. In the open cluster catalogue of \cite{hunt_improving_2024}, there are two clusters in roughly the same sky region at different distances: OC\,697 ($\varpi=0.397\pm0.045$~mas, distance $2282\pm12$~pc) and UFMG\,82 ($\varpi=0.59\pm0.05$~mas, distance $1550\pm9$~pc), both have Tr\,27 as an alternate designation. We were unable to find reliable radial velocities for the clusters. This could offer an explanation for the uncertain distance of Tr\,27 with results from different authors varying from 1000 to 2000~pc \citep{moffat_trumpler_1977}. The proper motion of V925\,Sco does not align with the members of either of these open clusters. It is in slightly better alignment with association no.~82, at a distance of 2250~pc \cite{chemel_search_2022}, close to OC\,697. We applied a $0.3$~{\masyr} proper motion cut around V925\,Sco's value on stars within 10{\arcmin} of the target. We found stellar groups at two distances: at $2310_{-110}^{+126}$~pc, the neighbourhood of OC\,697, and at $3110^{+359}_{-298}$~pc, a more distant population. V925\,Sco could belong to either population, but due to the proper motion similarity with the OB association we list the closer distance in Table~\ref{tab:distances}. Using the \textit{Gaia} DR3 radial velocity of -15.6~\kms, we found a \ion{H}{I}-based distance in the range of 1800 to 4000~pc. Because we are looking in the direction of the Galactic centre, the \ion{H}{I} velocity profile is rather flat around $-12$~\kms over a wide distance range. 

\textit{\object{[FMR2006] 15}}. The \textit{Gaia} DR3 parallax is low-quality ($\varpi/\sigma_\varpi=1.23$). This star is a member of the Red Supergiant Cluster~1 (RSGC~1) at a large distance of 5800~pc \citep{figer_discovery_2006} or $6600\pm890$~pc \citep{davies_cool_2008}. It is too distant to reliably detect a stellar group of hot stars. Using the radial velocity of $v_\mathrm{LSR}=120.8$~\kms \citep{davies_cool_2008} with an LSR correction of 15~\kms, we found a \ion{H}{I}-based distance of $\sim$6500~pc.

\textit{\object{IRAS 18357-0604}}. The \textit{Gaia} DR3 parallax is negative. This star has a very similar proper motion to nearby supergiants belonging to the cluster RSGC~2, at a distance of $5830^{1910}_{-780}$~pc \citep{davies_massive_2007}. Using the radial velocity of $90\pm3$~\kms \citep{clark_iras_2014}, we found a \ion{H}{I}-based distance of 5500~pc. 

\textit{\object{HD 179821}}. The \textit{Gaia} DR3 parallax is high-quality ($\varpi/\sigma_\varpi=9.2$), resulting in a distance of $4432^{+349}_{-355}$~pc \citep{bailer-jones_estimating_2021}. We found no alignment with any known cluster or OB association, but the star may be too distant. Using the radial velocity of $85.8\pm0.8$~\kms \citep{sahin_hd_2016}, we found a \ion{H}{I}-based distance of $>5000$~pc.  

\textit{\object{V1452 Aql}}. The \textit{Gaia} DR3 parallax is high-quality ($\varpi/\sigma_\varpi=15.5$), resulting in a distance of $2748^{+193}_{-148}$~pc \citep{bailer-jones_estimating_2021}. This star has been identified as a member of the open cluster CWNU\,1591 \citep{he_survey_2023}. We applied a $0.25$~{\masyr} proper motion cut around V1452\,Aql's value on stars within 10{\arcmin} of the target. We calculated a group distance of $2472^{+111}_{-104}$~pc. Using the \textit{Gaia} DR3 radial velocity of $33\pm8$~\kms, we found a \ion{H}{I}-based distance in the range of 2500--3500~pc.

\textit{\object{IRC +10420}}. The \textit{Gaia} DR3 parallax is medium-quality $(\varpi/\sigma_\varpi=3.43$), resulting in a distance of $4260^{+878}_{-752}$~pc. We did not found alignment with any known cluster or OB association, but the star may be too distant. Using the radial velocity of $\sim$60~{\kms} \citep{klochkova_optical_1997,jones_irc_1993,oudmaijer_spectral_1996}, we found a \ion{H}{I}-based distance of $\sim$5500~pc. This is in agreement with previous distance estimates in the range of 4000--6000~pc \citep{jones_irc_1993,reddy_spectroscopic_1999}.

\textit{\object{V1027 Cyg}}. The \textit{Gaia} DR3 parallax is is high-quality ($\varpi/\sigma_\varpi=13.45$), resulting in a distance of $3723^{+265}_{-237}$~pc. The star's proper motion aligns with the OB association no.~15 at a distance of 3902~pc \citep{chemel_search_2022}. We applied a $0.1$~{\masyr} proper motion cut around V1027\,Cyg's value on stars within 10{\arcmin} of the target. We calculated a group distance of $3977^{+226}_{-208}$~pc. Using the \textit{Gaia} DR3 radial velocity of $8.7\pm10.3$~{\kms}, we found a \ion{H}{I}-based distance in a very wide range of 2000--4500~pc. 

\textit{\object{HD 331777}}. The \textit{Gaia} DR3 parallax is high-quality ($\varpi/\sigma_\varpi=11.58$), resulting in a distance of $4559^{+388}_{-363}$~pc. On the sky, the star is located within the open cluster Kronberger~54 at a distance of $4025^{+73}_{-70}~pc$ \citep{hunt_improving_2024} and its proper motion aligns with cluster members within $3\sigma$. Its proper motion also aligns with members of the OB association no.~15 at a distance of 3902~pc \citep{chemel_search_2022}. We applied a $0.25$~{\masyr} proper motion cut around HD\,331777's value on stars within 10{\arcmin} of the target. We calculated a group distance of $3789^{+190}_{-176}$~pc. Using the \textit{Gaia} DR3 radial velocity of $0.8\pm5.5$~{\kms}, we found a \ion{H}{I}-based distance in a very wide range of 2000--5000~pc.

\textit{\object{RW Cep}}. The \textit{Gaia} DR3 parallax is medium-quality ($\varpi/\sigma_\varpi=3.33$), resulting in a distance of $6666^{+1561}_{-1006}$~pc. The star's proper motion aligns with the open cluster Berkeley~94 at a distance of $4000\pm40$~pc \citep{hunt_improving_2024} within $3\sigma$ and it has been identified as a cluster member \citep{delgado_berkeley_2013} or a member of the Cep\,OB1 association \citep{humphreys_studies_1978}. Its proper motion aligns with the OB association no.~122 \citep{chemel_search_2022} at a distance of 3876~pc. We applied a $0.1$~{\masyr} proper motion cut around RW\,Cep's value on stars within 10{\arcmin} of the target. We calculated a group distance of $3921^{+168}_{-157}$~pc. Using the radial velocity of $-50\pm3$~\kms \citep{kasikov_atmospheric_2025}, we found a \ion{H}{I}-based distance of $\sim$3500~pc.

\textit{V509\,Cas}. The \textit{Gaia} DR3 parallax is medium-quality ($\varpi/\sigma_\varpi=3.96$), resulting in a distance of $3917^{+969}_{-737}$~pc. It has been identified as a member of Cep\,OB1 association \citep{melnik_kinematics_2017,humphreys_studies_1978}. Its proper motion aligns with OB association no.~120 at a distance of 3055~pc \citep{chemel_search_2022}, which has proper motions consistent with the Cep\,OB1 association \citep{melnik_kinematics_2017}. We applied a $0.4$~{\masyr} proper motion cut around V509\,Cas' value on stars within 10{\arcmin} of the target. We calculated a group distance of $3368\pm127$~pc. This distance is about 2.5 times larger than the distance of $1370^{+561}_{-398}$ from the \textit{Hipparcos} parallax used by \cite{nieuwenhuijzen_hypergiant_2012}. Using the radial velocity of $-60.7$~{\kms} \citep{kasikov_yellow_2024},  we found a \ion{H}{I}-based distance of $\sim$3800~pc.

\textit{\object{6 Cas}}. The group distance determination for 6\,Cas is given in detail in Sect.~\ref{sect:dist_example}. Using the \textit{Gaia} DR3 radial velocity of $-43\pm8$~{\kms}, we found a \ion{H}{I}-based distance in the range of 2000--3000~pc. 

\textit{HD 223767}. The \textit{Gaia} DR3 parallax is high-quality ($\varpi/\sigma_\varpi=20.17$), resulting in a distance of $2879^{+136}_{-117}$~pc \citep{bailer-jones_estimating_2021}. It has been identified as a member of Cas\,OB5 association \citep{melnik_kinematics_2017,quintana_quantifying_2025}. On the sky, the star is very close to the open cluster King\,12 (distance $2805\pm19$~pc; \citealt{hunt_improving_2024}), but its proper motion is slightly outside of the 3$\sigma$ radius. Its proper motion aligns with the OB association no.~141 at a distance of distance 2827~pc  \citep{chemel_search_2022}. We applied a $0.2$~{\masyr} proper motion cut around HD\,223767's value on stars within 10{\arcmin} of the target. We calculated a group distance of $2740_{-83}^{+86}$~pc. Using the \textit{Gaia} DR3 radial velocity of $-41\pm2$~{\kms}, we found a \ion{H}{I}-based distance in the range of 2000--3000~pc. 

\textit{$\rho$ Cas}. The \textit{Gaia} DR3 parallax is negative. The star has been identified as a member of the Cas OB5 association \citep{melnik_kinematics_2017}. However, we found that the proper motions of the association members differ by $\sim$1~mas~yr$^{-1}$ and the association is located >2 deg away on the sky. We did not find alignment with any known cluster or OB association. We applied a $1$~{\masyr} proper motion cut around $\rho$\,Cas' value on stars within 20{\arcmin} of the target. As there are very few hot stars in the region, we included cooler stars down to 7000~K. Among stars with similar proper motions appear two distinct groups: six stars at 1700--2100~pc (\textit{Gaia} parallaxes between 0.47 and 0.54) and 17 stars at 2500--3500~pc (parallaxes between 0.27 and 0.38). The second group includes three stars with \teff > 8000~K, all other stars in both groups are cooler. For estimating the distance of $\rho$\,Cas, we used the mean parallax of the second group. We calculated a group distance of $2810_{-102}^{+104}$~pc. Using the radial velocity of $-47\pm2$~{\kms} \citep{lobel_high-resolution_2003}, we found a \ion{H}{I}-based distance in the range of 2500--3200~pc, in agreement with the distance to the farther stellar group. 

\clearpage
\onecolumn
\section{Proper motion cuts for group distance determination}\label{app:prop_mot_fig}
\begin{figure}[htb!]
    \centering
    \includegraphics[width=\linewidth]{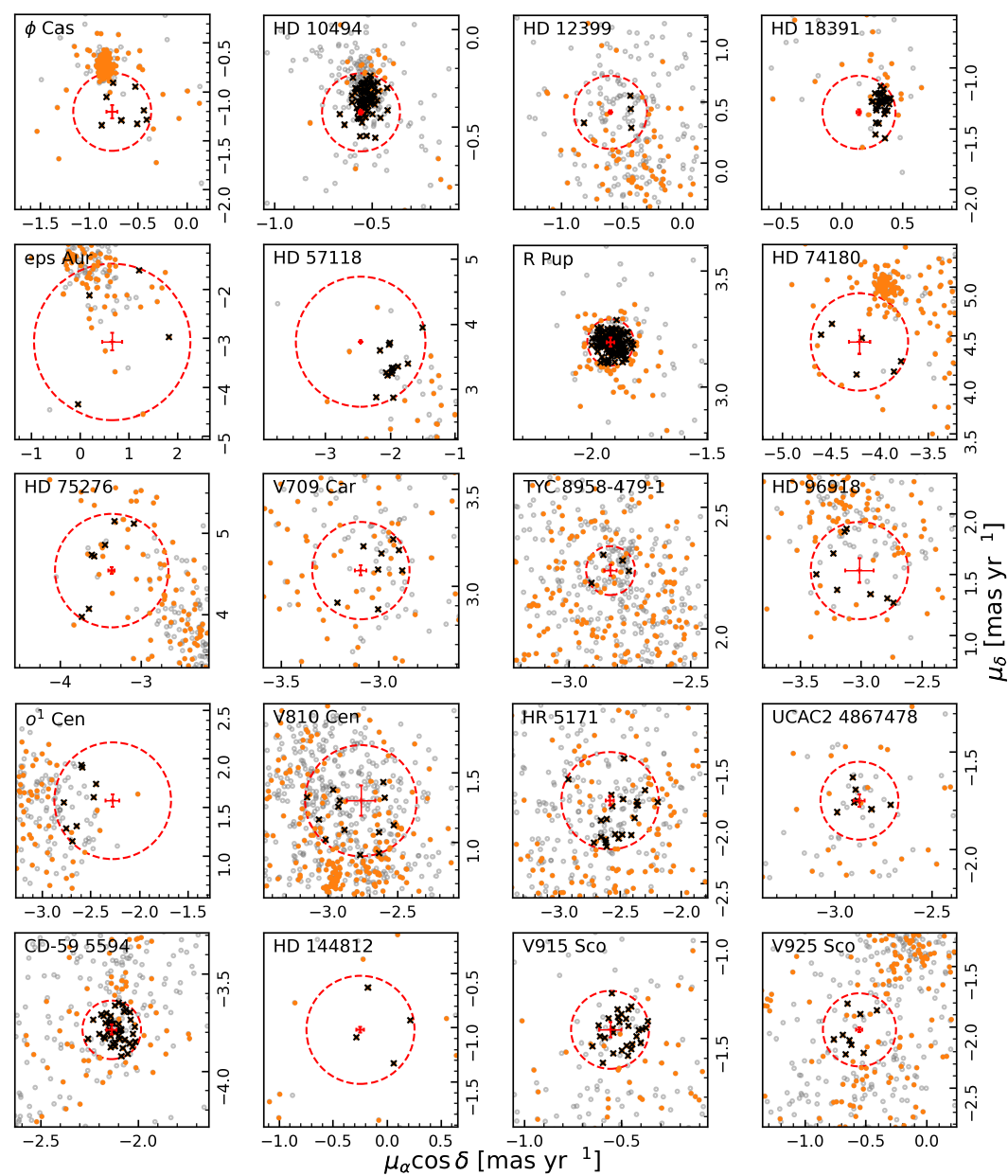}
    \caption{Selection of hot stars around each sample YSG/YHG based on proper motion criteria. The YSG/YHG is in the centre with error bars. The red dashed circle marks the proper motion cut radius. Stars shown as orange circles meet the selection criteria from Sect.~\ref{sect:distance}. Stars used for calculating the group distance are marked with black crosses.}
    \label{appfig:prop_mot_fig_1}
\end{figure}

\begin{figure}[h!]
    \centering
    \includegraphics[width=\linewidth]{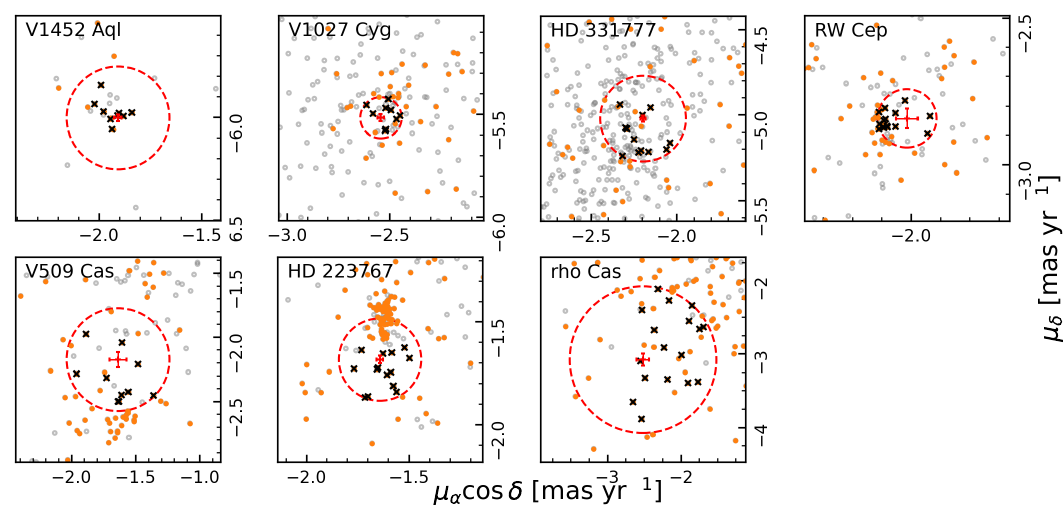}
    \caption{Fig.~\ref{app:prop_mot_fig}.1. continued.}
    \label{appfig:prop_mot_fig_2}
\end{figure}

\section{Stars selected for group distance determination}\label{apptable:nearby_stars_info}

\begin{table}[h!]
\caption{Stars selected for the group-based distance calculation. Listed are the relevant \textit{Gaia} DR3 parameters used for proper motion bias and parallax zero-point corrections. The distance from \cite{bailer-jones_estimating_2021} is also included. The table includes the corrected proper motions, the corrected parallaxes and external parallax uncertainties. Shown here are the first three rows for $\phi$\,Cas. }
\centering
\begin{tabular}{lllllllll}
\hline\hline
\multicolumn{9}{l}{$\phi$\,Cas} \\
\hline
SOURCE\_ID                    & ra           & dec      & parallax                    & parallax\_error    &  &  &  &  \\ \hline
413874460384818432            & 19.929       & 58.243   & 0.396                       & 0.018              &  &  &  &  \\
413827043942817792            & 19.992       & 58.172   & 0.361                       & 0.022              &  &  &  &  \\
413876556328777728            & 19.972       & 58.333   & 0.29                        & 0.02               &  &  &  &  \\
...                           &              &          &                             &                    &  &  &  &  \\ \hline
pmra                          & pmra\_error  & pmdec    & pmdec\_error                & phot\_g\_mean\_mag &  &  &  &  \\\hline
-0.966                        & 0.013        & -1.33    & 0.014                       & 14.018             &  &  &  &  \\
-0.835                        & 0.017        & -1.191   & 0.021                       & 14.973             &  &  &  &  \\
-1.277                        & 0.014        & -1.292   & 0.016                       & 14.645             &  &  &  &  \\
\multicolumn{9}{l}{...}  \\ \hline
nu\_eff\_used\_in\_astrometry & pseudocolour & ecl\_lat & astrometric\_params\_solved &                    &  &  &  &  \\ \hline
1.544                         & nan          & 45.135   & 31                          &                    &  &  &  &  \\
1.529                         & nan          & 45.057   & 31                          &                    &  &  &  &  \\
1.574                         & nan          & 45.199   & 31                          &                    &  &  &  &  \\
\multicolumn{9}{l}{...}  \\ \hline
pmra\_corr                    & pmdec\_corr  & rgeo     & b\_rgeo                     & B\_rgeo            &  &  &  &  \\ \hline
-0.966                        & -1.33        & 2353.305 & 2251.713                    & 2445.136           &  &  &  &  \\
-0.835                        & -1.191       & 2570.845 & 2431.047                    & 2744.906           &  &  &  &  \\
-1.277                        & -1.292       & 3090.492 & 2907.08                     & 3318.825           &  &  &  &  \\
\multicolumn{9}{l}{...} \\ \hline
zpt                           & plx\_corr    & k\_value & plx\_ext\_uncert            & weight             &  &  &  &  \\ \hline
-0.032                        & 0.428        & 1.498    & 0.028                       & 0.161              &  &  &  &  \\
-0.032                        & 0.392        & 1.403    & 0.033                       & 0.121              &  &  &  &  \\
-0.032                        & 0.322        & 1.436    & 0.03                        & 0.144              &  &  &  &  \\
\multicolumn{9}{l}{...} \\  \hline
\end{tabular}
\end{table}
\twocolumn

\clearpage
\section{Radial velocities}
\begin{table}[H]
\caption{\label{app:lit_rvs} Radial velocities compiled from the literature. For \textit{Gaia} DR3 values, the errors are 1/2 of the total radial velocity variability amplitude. }
\begin{tabular}{l|p{18mm}p{33mm}}
\hline\hline
Identifier & $v_\mathrm{rad}$~(\mbox{\kms}) & Ref. \\ \hline
$\phi$\,Cas & \mbox{$-19.2\pm3.0$} \newline \mbox{$-28 \pm 3$} \newline \mbox{$-28.5\pm1.2$} & \textit{Gaia} DR3 \newline \cite{arellano_ferro_comments_1988} \newline \cite{rosenzweig_determination_1993} \\
HD\,10494 & \mbox{$-30.6\pm1.4$} \newline $-35\pm1$ & \textit{Gaia} DR3 \newline \cite{smolinski_radial_1980} \\
HD\,12399 & \mbox{$-44.1\pm7.6$} & \textit{Gaia} DR3 \\
HD\,18391 & \mbox{$-38.5\pm2.8$} & \textit{Gaia} DR3 \\
$\epsilon$ Aur & \mbox{$-2.26 \pm 0.15$} & \cite{stefanik_epsilon_2010} \\
HD\,57118 & \mbox{$61.2\pm1.4$} & \textit{Gaia} DR3 \\
R\,Pup & \mbox{$69.1\pm4.0$} \newline \mbox{$62.5 \pm 0.5$} & \textit{Gaia} DR3 \newline \cite{balona_radial_1982} \\
HD\,74180 & \mbox{$25\pm3$} & \cite{forbes_pismis_1994} \\
HD\,75276 & \mbox{$26.7\pm1.2$} \newline $32.0\pm2.5$ \newline $21.6\pm0.47$ & \textit{Gaia} DR3 \newline \cite{reed_vela_2000} \newline \cite{de_medeiros_catalog_2002} \\
V709\,Car & \mbox{$8.4\pm8.0$} \newline \mbox{$12\pm6$} & \textit{Gaia} DR3 \newline \cite{maas_study_2003} \\
HD\,96918 & \mbox{$5.6 \pm 0.5$} \newline 7.3 & \cite{balona_radial_1982} \newline \cite{humphreys_studies_1978} \\
$o^1$\,Cen & \mbox{$-21.1\pm9.5$} \newline $-20.8\pm0.6$ & \textit{Gaia} DR3 \newline \cite{gontcharov_pulkovo_2006} \\
V810\,Cen & \mbox{$-16.7 \pm 3.7$} & \cite{kienzle_pulsating_1998} \\
HR\,5171A & $-40$ \newline $-38$ & \cite{humphreys_high-luminosity_1971} \newline \cite{balona_radial_1982} \\
UCAC2\,4867478 & \mbox{$-41.4\pm3.6$} & \textit{Gaia} DR3 \\
CD-59\,5594 & \mbox{$-35.7\pm4.2$} & \textit{Gaia} DR3 \\
V870\,Sco & \mbox{$-43.2\pm5.5$} \newline $-53$ & \textit{Gaia} DR3 \newline \cite{herbig_highly_1972} \\
V915\,Sco & $-6$ & \cite{andrews_hr_1977} \\
IRAS\,17163-3907 & $18$* & \cite{wallstrom_alma_2017} \\
V925\,Sco & \mbox{$-15.6\pm5.8$} \newline \mbox{$-14.7\pm1.7$} & \textit{Gaia} DR3 \newline \cite{kipper_optical_2008} \\
{[FMR2006]}\,15 & 102.2* \newline 120.8* & \cite{figer_discovery_2006} \newline \cite{davies_cool_2008} \\
IRAS\,18357-0604 & $90\pm3$ & \cite{clark_iras_2014} \\
HD\,179821 & \mbox{$85.8\pm0.8$} & \cite{sahin_hd_2016} \\
V1452\,Aql & \mbox{$32.7\pm8.3$} \newline $32\pm1.5$ & \textit{Gaia} DR3 \newline \cite{smolinski_radial_1980} \\
IRC\,+10420 & \mbox{$60-68$} \newline \mbox{$76.2\pm1.1$*} \newline $77$* & \cite{klochkova_optical_1997} \newline \cite{jones_irc_1993} \newline \cite{oudmaijer_spectral_1996} \\
V1027\,Cyg & \mbox{$8.7\pm 10.3$} \newline 5.5 & \textit{Gaia} DR3 \newline \cite{klochkova_new_2016} \\
HD\,331777 & \mbox{$0.8\pm5.5$} & \textit{Gaia} DR3 \\
RW\,Cep & \mbox{$-50.3\pm3.3$} \newline \mbox{$-53.3\pm4.0$} & \cite{kasikov_atmospheric_2025} \newline \textit{Gaia} DR3 \\
V509\,Cas & $-60.7$ & \cite{kasikov_yellow_2024} \\
6\,Cas & \mbox{$-43.0\pm8.4$} & \textit{Gaia} DR3 \\
HD\,223767 & \mbox{$-41.2\pm2.1$} & \textit{Gaia} DR3 \\
$\rho$\,Cas & \mbox{$-47\pm2$} & \cite{lobel_high-resolution_2003}\\
\hline
\end{tabular}\\
* $v_\mathrm{LSR}$ velocity
\end{table}

\end{appendix}
\end{document}